\documentclass[a4paper,fleqn,usenatbib]{mnras}

\usepackage[T1]{fontenc}
\usepackage{ae,aecompl} 
\usepackage{threeparttable}

\usepackage{graphicx}	
\usepackage{amsmath}	
\usepackage{amssymb}	
\usepackage{txfonts}
\usepackage[normalem]{ulem}  
\usepackage{color}
\usepackage{natbib}
\bibpunct[; ]{(}{)}{,}{a}{}{,}

\def\widerul{\vrule height 2.5ex width 0ex depth 0ex}

\def\asec{\ifmmode ^{\prime\prime}\else$^{\prime\prime}$\fi}

\def\it{\sl}
\def\degs{\ifmmode ^{\circ}\else$^{\circ}$\fi}
\def\amin{\ifmmode ^{\prime}\else$^{\prime}$\fi}
\def\asec{\ifmmode ^{\prime\prime}\else$^{\prime\prime}$\fi}
\def\farcs{\hbox{$.\!\!^{\prime\prime}$}}  

\def\psr{PSR~B0656+14\ }
\def\degs{\ifmmode ^{\circ}\else$^{\circ}$\fi}
\def\amin{\ifmmode ^{\prime}\else$^{\prime}$\fi}
\def\farcm{\hbox{$.\mkern-4mu^\prime$}}
\def\eqalign#1{\null\,\vcenter{\openup1\jot \m@th
   \ialign{\strut\hfil$\displaystyle{##}$&$\displaystyle{{}##}$\hfil
   \crcr#1\crcr}}\,}

\title[ PSR B0656+14]{PSR B0656+14: the unified outlook from the  infrared to X-rays}

\author[S. Zharikov, D. Zyuzin,  Yu. Shibanov,  et al.]{
S. Zharikov,$^{1}$\thanks{E-mail: zhar@astro.unam.mx}
D. Zyuzin,$^{2}$
 Yu. Shibanov,$^{2}$
A. Kirichenko,$^{1,2}$
R. E. Mennickent,$^{3}$
 S. Geier,$^{4,5}$ 
\newauthor A. Cabrera-Lavers$^{4,5}$
\\
$^{1}$ Instituto de Astronom\'ia, Universidad Nacional Aut\'onoma de M\'exico, Apdo. Postal 877, Ensenada, Baja California, M\'exico, 22800\\
$^{2}$ Ioffe Institute, 26 Politekhnicheskaya st., St. Petersburg 194021, Russia\\
$^{3}$ Departamento de Astronom\'{i}a, Universidad de Concepci\'on, Casilla 160-C, Concepci\'{o}n, Chile \\
$^4$ Instituto de Astrof\'isica de Canarias, V\'ia L\'actea s/n, E38200, La Laguna, Tenerife, Spain\\
$^5$ GRANTECAN, Cuesta de San Jos\'e s/n, E-38712, Bre\~{n}a Baja, La Palma, Spain
}

\date{Accepted XXX. Received YYY; in original form ZZZ}

\pubyear{2019}

\begin{document}
\label{firstpage}
\pagerange{\pageref{firstpage}--\pageref{lastpage}}
\maketitle

\begin{abstract} 

We report  detection of PSR~B0656$+$14 with the Gran Telescopio Canarias in  narrow optical $F657$, $F754$, $F802$, and $F902$  
and  near-infrared $JHK_s$ bands.
The pulsar detection in the  $K_s$ band extends its spectrum to 2.2 $\mu$m and confirms its flux increase towards the  infrared.
We also present a thorough analysis of the optical
spectrum obtained by us with the VLT. For a consistency check, we revised the pulsar near-infrared and narrow-band photometry 
obtained with the \textit{HST}. We find no
narrow spectral lines in the optical spectrum. We compile  available near-infrared-optical-UV and  archival 0.3-20~keV  
X-ray data and perform a self-consistent
analysis of the rotation phase-integrated spectrum of the pulsar using unified spectral models. The spectrum is best fitted 
by the four-component model including two
blackbodies, describing the thermal emission from the neutron star surface and its hot polar cap, the broken power-law, 
originating from the pulsar magnetosphere, and an
absorption line near $\sim$0.5 keV detected previously. The fit provides better constraints on the model parameters 
than using only a single spectral domain. The derived
surface temperature is $T_{NS}^{\infty}=7.9(3)\times10^5$~K. The intrinsic radius  (7.8-9.9~km) of the emitting region 
is smaller than a typical neutron star radius (13~km)
and suggests a nonuniform temperature distribution over the star surface. In contrast, the derived radius of the hot polar 
cap is about twice as large as the `canonical' one.
The spectrum of the nonthermal emission steepens from the optical to X-rays and has a break near 0.1~keV. The X-ray data
suggest the presence of another absorption 
line near 0.3~keV.

\end{abstract}



\section{Introduction}
\label{intro}

More than fifty years after the discovery of the first neutron stars (NSs) as radio pulsars,
their emission mechanisms still remain poorly understood, and multiwavelength observations are
crucial for any progress in this field. Pulsars produce emission across
the entire electromagnetic spectrum, being mostly observed as powerful radio, X-ray and
$\gamma$-ray emitters. In contrast, their detection in the optical and/or near-infrared 
(near-IR) is still a rare event. The studies of most handful of them show that the optical
emission consists of the thermal spectral component from the surface of the NS and the 
nonthermal component from its magnetosphere. Contribution of the  components depends on 
the pulsar age. In turn, the near-IR emission is believed to have a purely magnetosphere nature.

The first optical and near-IR identifications of NSs have been reported shortly after the discovery
of radio pulsars \citep{cocke} and to date number about 25 firm detections (see, e.g., 
\citet{2011AdSpR..47.1281M} for review). However, despite the rapidly increasing amount 
of known pulsars\footnote{http://www.atnf.csiro.au/people/pulsar/psrcat/, \citet{manchester}},  
no significant progress has been made in the field during the last decade. Great expectations
were set on the discoveries of the \textit{Fermi} Large Area Telescope, 
which was launched in 2008 and has detected more than 200 $\gamma$-ray pulsars. Most of them
appeared to be nearby and energetic, making them promising targets for 
optical and near-IR studies and boosting searches with the Hubble Space Telescope ({\it HST}) and the
largest ground-based telescopes. 
However, among numerous optical and near-IR observations of $\gamma$-ray pulsars, only three  
possible identifications were proposed \citep{zyuzin, mignani16a, rangelov}. Nevertheless, 
deep upper limits on the optical fluxes of other pulsars 
provided informative constraints on their emission properties and multiwavelength spectra (see,
e.g., \citet{2013MNRAS.435.2227Z}, \citet{mignani2018} and references therein).

Most of the identified NS optical counterparts  are very faint ($V \geq 25$) and can be  
detected only via broad-band photometric observations.    
Optical spectroscopy has been obtained only for the youngest and brightest Crab pulsar 
($\sim$1000 yr., $V=16.6$) \citep{1996A&A...314..849N, 2000ApJ...537..861S, 2001A&A...374..584B}, 
PSR B0540$-$69 ($\sim$2000 yr., $V=22.4$) \citep{1997ApJ...486L..99H, 2004A&A...425.1041S} and 
the third  optically brightest Vela pulsar ($\sim$10000 yr., $V=23.6$)
\citep{2007A&A...473..891M}. The spectroscopic observations of older and fainter pulsars were
first obtained by \citet{1998ApJ...494L.211M}. They reported observations of the 
Geminga pulsar ($\sim$ $3\times10^5$ yr., $R=25.5$) using the Keck telescope.  
The preliminary optical spectroscopy for another middle-aged \psr\ ($\sim$10$^5$ yr., $R=24.6$)
was reported by \citet{2007Ap&SS.308..545Z}.  
In addition, spectroscopic observations of the isolated radio-silent neutron star (INS) 
RX~J1856.5-3754 ($\sim$ $5\times10^5$ yr, $V=25.6$) 
were performed with the Very Large Telescope (VLT) by \citet{vlt-rxj1856}.
The young Crab,  Vela, and middle-aged Geminga  pulsars demonstrate an almost a 
flat spectrum in the optical and near-IR \citep{Crab2009,2006A&A...448..313S, Zyuzin:2013aa}.
 In contrast, young B0540$-$69 \citep{2019ApJ...871..246M} and the middle-aged B0656+14 \citep{2006A&A...448..313S}
exhibit a flux increase towards the near-IR.
Moreover, the optical-IR data of  B0656+14 suggests the presence of 
a spectral break between the two ranges. 
Thorough studies in the near-IR are crucial to
broaden the distribution and reveal the presence of similar features in other objects, which is
important for understanding the physics of emitting particles in pulsar magnetospheres.

\begin{table*}
\caption{ Parameters of \psr obtained from radio observations \citep{1993ApJS...88..529T}.
The pulsar proper motion and distance are adopted from \citet{Brisken:2003aa}.}
\label{t:656_prop}
\begin{tabular}{cccccccccccc}
\hline\hline
\multicolumn{7}{c}{Observed}&&\multicolumn{4}{c}{Derived} \widerul\\
\cline{1-8}\cline{10-12}
$P$        & $\dot P$   & $D\!M$       & $l$   & $b$ & $\mu_\alpha$  & $\mu_\delta$   & $d$  && $\tau$ & $B$                   & $\dot E$                \widerul \\
ms         & $10^{-14}$ & cm$^{-3}$ pc & deg   & deg & mas yr$^{-1}$   & mas yr$^{-1}$   & pc  && Myr    & G                     & erg $s^{-1}$               \\
\hline
384.87     & 5.50       & 13.977         & 201.1 & 8.3 & $44.07\pm0.63$        & $-2.40\pm.29$  & $288^{+33}_{-27}$      && 0.11   & $4.7 \times 10^{12}$  & $3.8 \times 10^{34}$   \widerul \\
\hline
\end{tabular}
\begin{tabular}{c}
   $P$ - spin period; $\dot{P}$ - derivative of the spin period;   $DM$- the dispersion measure; $l,b$-  Galactic coordinate;  $\mu_\alpha$,  $\mu_\delta$- proper motion; \\
   $d$- distance; $\tau$  - the pulsar age; $B$- magnetic field strength; $\dot{E}$- spin-down luminosity
     
\end{tabular}
\label{t:1}
\end{table*}

Owing to its relative brightness and proximity, 
the middle-aged \psr is one of those isolated NSs most intensively studied in different
wavelengths. It was first discovered as a radio pulsar by \citet{1978MNRAS.185..409M}.
The average pulse flux densities of the pulsar, 6.5(6), 3.8(6), 4.8(1.4) and 3.7(8) mJy
at 0.4, 0.6, 0.9 and 1.4 GHz, respectively, result in a power-law (PL) spectrum  
$S\sim \nu^{-\alpha}$ with the index $\alpha=0.5(2)$ \citep{Lorimer:1995aa}. 
The \psr pulse profile is almost completely linearly 
polarised with a shallow asymmetric swing of the position angle \citep{Gould:1998aa}. 
\citet{Kuzmin:2006aa} suggested and \citet{Weltevrede:2006aa} and \citet{Tao:2012aa} 
confirmed that \psr belongs to the group of pulsars that emit giant pulses with the peak flux
density of the strongest pulse by a factor of 630 higher than of the average pulse. 
The pulsar dispersion measure is 13.977 pc cm$^{-3}$. 
\citet{Brisken:2003aa} measured the parallax $\upi=3.47(36)$~mas with the Very Long Baseline Array 
yielding a distance of 288$^{+33}_{-27}$ pc. 
\psr is located near the centre of the Monogem ring \citep{Nousek:1981aa}, a $\sim10^5$ yr
expanding supernova remnant (SNR) 
likely produced in the same supernova explosion as the pulsar \citep{Thorsett:2003aa}.
 The basic parameters of \psr inferred from the radio observations are given in Table~\ref{t:1}. 
 
\begin{figure*}
\setlength{\unitlength}{1mm}
\resizebox{15.cm}{!}{
\begin{picture}(130,75)(0,0)
\put (-10,0){{\includegraphics[width=15.cm, clip=]{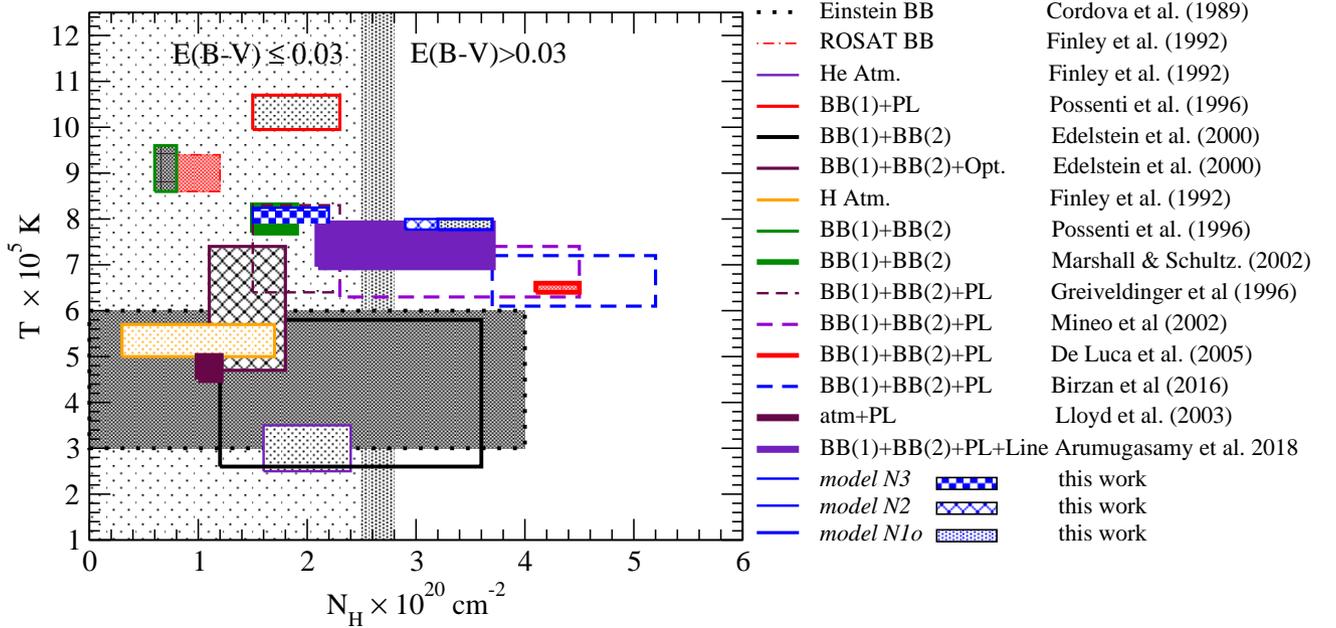}}}
\end{picture}}

\caption{Diagram of the surface temperature $T$ of the NS {\sl vs} the absorbing column density 
N$_H$ towards the pulsar. These are compiled results of various  spectral fits of \psr\  X-ray emission 
published by different authors listed in the legend. In the legend, BB(1) is the low-temperature blackbody 
spectral component from the bulk of the NS surface,  BB(2) is the high-temperature blackbody component from 
the hot polar spots of the pulsar and  PL corresponds to the nonthermal component originating from the pulsar
magnetosphere. Boxes show various 1$\sigma$   constraints on  the BB(1) temperature and  N$_H$ taken from
the publications (see Table~\ref{tab:X-ray}). The result of this work is also announced.  The most plausible 
for the pulsar colour-excess range $E(B-V)\le0.03$ is shown by the short-dash-hatched region  
(see text for details).   }
\label{FigX-rayFits}
\end{figure*}

The pulsar has been detected and intensively studied in X-rays with the space observatories
{\it Einstein} \citep{1989ApJ...345..451C}, {\it ROSAT} \citep{1992ApJ...394L..21F,
1993ApJ...414..867A, possenti},  {\it ASCA} \citep{1996ApJ...465L..35G}, {\it BeppoSAX}
\citep{mineo}, {\it XMM-Newton} \citep{2004MmSAI..75..458Z, De-Luca:2005aa}, {\it Chandra}
\citep{2002ApJ...574..377M, Pavlov:2002aa, PA-model2003,  Birzan:2016aa} and {\it NuSTAR}  
\citep{ 2018XMM}. It has also been observed with 
\textit{ SRG/eROSITA} during the verification stage at the end of
2019\footnote{http://www.mpe.mpg.de/7362694/presskit-erosita-firstlight}. 
The observed flux in the 0.2-3.0 keV band is about $1.01(3)\times10^{-11}$ 
ergs cm$^{-2}$ s$^{-1}$.
 The observed X-ray spectrum consists of at least two distinct components, thermal and non-thermal.
It appeared to be best fitted 
by an absorbed   blackbody (BB) 
model with a relatively low temperature T~$\approx~(6 - 7) \times 10^5$~K combined with 
the additional  BB  with a higher temperature of $\approx~1.3 \times 10^6$ K  and the PL
describing the  hard X-ray tail \citep{De-Luca:2005aa}.
The non-thermal PL component of the pulsar magnetosphere origin contributes only 
1\% of the total luminosity in the 0.1-10 keV range. The low temperature BB component represents
the thermal emission from  the bulk of the  surface of the NS, while the hot BB component  
is from its small  polar caps heated by relativistic particles  from the magnetosphere 
of the pulsar.  
A historical  summary of the X-ray spectral fits including the {\it EUVE} Deep Survey (DS) data
\citep{2000ApJ...539..902E} 
is presented in Fig.~\ref{FigX-rayFits} and Table~\ref{tab:X-ray}.
The temperature of the bulk surface of the pulsar provided by the cold BB component  despite of a noticeable uncertainty
is consistent with prediction of cooling scenarios  of  NSs \citep{2015MNRAS.447.1598B}.  
However, the 2BB+PL model gives a strong scatter in the absorbing column density 
N$_{\mathrm H}=(1.5-5.5)\times10^{20}$ cm$^{-2}$. 
The X-ray observations also showed 
a complex behavior of the source pulsed fraction: it decreases with 
energy from about 20\% at $\sim$0.1-0.3keV down to about 7-11\% at around 1 keV and increases 
again to $\approx$35-44\% at $\gtrsim$1 KeV \citep{Pavlov:2002aa, PA-model2003, De-Luca:2005aa}.
Recently, \citet{2018XMM} reported 
the spectral analysis of new data obtained with 
\textit{XMM-Newton} and \textit{NuSTAR}.  
They found that the 2BB$+$PL model 
is not consistent with the  new data due to large fit-residuals in the 0.3--0.7  keV range.       
  However, the residuals can be naturally explained 
by the presence of an absorption line in the spectrum of the pulsar. The latter has been 
confirmed at a higher confidence by the recent \textit{ SRG/eROSITA} data \citep{eRosita}.
Using  NS atmosphere models, such as \textit{NSA} \citep{1996A&A...315..141Z} and \textit{NSMAXG}  
\citep{2014IAUS..302..435H}, 
does not lead to  acceptable fits either. 
  
\begin{table*}
\caption{Summary of X-ray spectral analyses of \psr published  during the period 1989--2018.   }
\begin{tabular}{lcllcl}
\hline\hline
Satellite   & Range & Models                                &  Temperature/spec. index   & N$_H$     & Reference  \\
                & keV   &   (1)-soft; (2)-hard                                          &    \hspace{0.05cm}  $\times10^5$ K / $F\propto \nu^\alpha \propto \nu^{-\Gamma +1}$                                      & cm$^{-2}$ &                  \\ \hline
{\it Einstein}   & 0.15 --- 4.0 keV & single BB(1);              & T$_1$= 3.0---6.0                                      & $<4 \times10^{20} $   &  \citet{1989ApJ...345..451C}  \\
	        & 0.15 --- 4.0 keV &                       PL         &                                                                &                                    &                                                 \\
{\it ROSAT}    & 0.1 --- 2.4 keV & single BB(1)                  &T$_1$= 8.6---9.4                                        & (0.8-1.2)$\times10^{20}$ &\citet{1992ApJ...394L..21F} \\
	       & 0.1 --- 2.4 keV & PL                                 & $\alpha_{PL}=-4.8$                                                           & (0.8-1.2)$\times10^{20}$ &  \\
	       & 0.1 --- 2.4 keV & He  Atm.                        &T$_1$= 2.5---3.5                                      & (1.6- 2.4)$\times10^{20}$ & \\
	       & 0.1 --- 2.4 keV & H Atm.                           &T$_1$= 5.0---5.7                                       & (0.3-1.7)$\times10^{20}$ &  \citet{1993ApJ...414..867A}\\
             & 0.1 --- 2.4 keV & BB(1)+BB(2)                 &T$_1$= 8.6---9.6                                      &  (0.6-0.8)$\times10^{20}$  &  \citet{possenti} \\
	       &                         &                                       &T$_2$= 15---23                                       &                                                        &                                                  \\
               & 0.1 --- 2.4 keV & BB(1)+PL                      &T$_1$= 9.95---1.07                                   &  (1.5-2.3)$\times10^{20}$  &  \citet{possenti} \\
	       &                        &                                       &$\alpha_{PL}=-3.4(4)$                                                         &                                                         &                                                   \\
{\it ASCA}     & 0.5 --- 10.0 keV & BB(1)+BB(2)+PL            &T$_1$= 6.4---8.3                              &  (1.5-2.3)$\times10^{20}$  &\citet{1996ApJ...465L..35G}  \\
	       &                        &                                       &T$_2$= 13---17                                        &                                                         &                                                   \\
	       &                        &                                       &$\alpha_{PL}=-0.5(1.1)$                                                 &                                                         &                                                   \\
{\it BeppoSAX}       & 0.1 --- 10.0 keV & BB(1)+BB(2)+PL         &T$_1$= 6.3---7.4                                   &  (2.3-4.5)$\times10^{20}$  & \citet{mineo} \\
	       &                        &                                       &T$_2$= 13---15                                                   &                                                         &                                                   \\
	       &                        &                                       &$\alpha_{PL}=-1.08(41)$                                                 &                                                         &                                                   \\
{\it Chandra} 
& 0.1 --- 2.0 keV & BB(1)+BB(2)                &T$_1$= 7.7---8.3                                   & (1.5-1.9)$\times10^{20}$  & \citet{2002ApJ...574..377M}\\
	       &                        &                                        &T$_2$= 13---19                                                 &                                                         &                                                   \\
{\it Chandra} 
& 0.3 --- 8.0 keV & BB(1)+BB(2)+PL          &T$_1$= 6.1---7.2                                  &  (3.7-5.2)$\times10^{20}$  &\citet{Birzan:2016aa}\\
	       &                        &                                        &T$_2$= 12.1---13.6                                                  &         {\sc wabs}                                                &          \\
               &                        &                                       &$\alpha_{PL}=-1.3(7)$                                                 &                                                         &                                       \\           
{\it XMM-Newton}       & 0.1 --- 10.0 keV & BB(1)+BB(2)+PL          &T$_1$= 6.4---6.6                                  &  (4.1-4.5)$\times10^{20}$  &         \citet{De-Luca:2005aa}     \\
	       &                        &                                       &T$_2$= 12.2---12.8                                    &                                                         &                                                   \\
	       &                        &                                       &$\alpha_{PL}=-1.1(3)$                                                 &                                                         &                                                   \\ 
{\it EUVE $+$ ROSAT}     &  0.1 --- 2.4 keV & BB(1)+BB(2)          &T$_1$= 2.6---5.8                                   &  (1.2-3.6)$\times10^{20}$  &         \citet{2000ApJ...539..902E}     \\
	       &                        &                                       &T$_2$= 10.6---12.8                                   &                                                         &                                                   \\
{\it EUVE $+$ ROSAT}     &  0.1 --- 2.4 keV & BB(1)+BB(2)          &T$_1$= 4.7---7.4                                   &  (1.1-1.8)$\times10^{20}$  &       \citet{2000ApJ...539..902E}      \\
+ Optical        &                        &                                       &T$_2$= 11---18                                   &                                              &                                                   \\ 
 {\it Chandra} $+$ Optical  
 &  0.3 --- 6 keV 
 & H Atm. + PL        & T= 4.8---6.2                                   &  (1.0-1.2)$\times10^{20}$  &       \citet{PA-model2003}      \\

                            & 
&                        & T$^{\infty}$=4.5---5                                    &                                              &                                \\   
                                        &                                       &                        & $\alpha_{PL}$ = -0.5                                  &                                              &                                      \\
{\it XMM} $+${\it NuStar}                 &  0.3 --- 7.0KeV    
 &   BB(1)+BB(2)+PL+Line   &      T= 6.8---7.9       &         (2.1-3.7)$\times10^{20}$                                     &         \citet{2018XMM}                     \\  
                                          &                                       &                        & $\alpha_{PL}$ = -0.7(1)                                  &                                              &                     ({\it A18})                 \\                      
  \hline
\end{tabular}
\begin{tabular}{l}
Comments:
BB(1) is the blackbody from the NS surface; BB(2) -- blackbody from its hot spot;  
PL -- power law component; He Atm -- He atmosphere \\ model; H Atm - H-atmosphere model.   \\ 
\end{tabular}
\label{tab:X-ray}
\end{table*}

\begin{table}
\caption{Log of the GTC observations of PSR B0656+14.}
\begin{tabular}{lccccl}
\hline\hline
 Date                   & Band                                                                    &  Exposure                     &  AM             &  Seeing &Observing    \\
   &    $\lambda$/$\Delta\lambda$                  & N$_{exp}\times$ [sec]   &  $\sec$(Z)                   &   [arcsec]            &   conditions             \\    \hline \hline                 
   2009-10-17      &   $F657/35$      &  6 $\times$ 250.0 &1.18    &   0.7   &   Photom.          \\
   2009-10-17      &   $F754/50$      &  9 $\times$ 200.0 &1.10    &  0.7    &   Photom.           \\  
   2011-10-28      &   $F902/44$      &  2 $\times$ 600.0 &1.03    &   0.7    &   Photom.          \\
   2011-10-28      &   $F902/44$      &  2 $\times$ 300.0 &1.04    &   0.7    &   Photom.                   \\   
   2011-10-29      &   $F902/44$      & 6 $\times$ 325.0 &1.04     &    0.7    &   Photom.                  \\
   2011-10-29      &   $F902/44$      & 6 $\times$ 325.0 &1.04     &    0.7    &   Photom.                   \\
   2011-11-21      &   $F902/44$      & 13 $\times$ 325.0 &1.03   &     0.7    &   Photom.                  \\ 
   2011-12-27      &   $F657/35$      & 2 $\times$ 300.0 &1.05     &   0.9    &     Photom.              \\
   2011-12-27      &   $F754/50$      & 2 $\times$ 300.0 &1.10     &   0.9    &   Photom.          \\
   2011-12-27      &   $F802/51$      & 12 $\times$ 300.0&1.12    &    0.9    &   Photom.          \\ 
   2012-01-23      &   $F902/44$      & 18 $\times$ 325.0&1.12    &     0.7    &   Photom.            \\  \hline
   2016-10-22      &  $K_s$        & 168~$\times$~10         & 1.21          &  0.8--1.6       & Photom.     \\
   2016-10-30      &  $J$          & 154~$\times$~30           & 1.05          &  0.8--1.2       & Clear                  \\
   2016-10-31      &  $H$          & 112~$\times$~20           & 1.04          &  0.5--0.8       & Clear                  \\
   2016-11-01      &  $H$          & 112~$\times$~20           & 1.04          &  0.5--0.7       & Clear                   \\
   2016-11-01      &  $J$          & 42~$\times$~30            & 1.08          &  0.5--0.6       & Clear                  \\
   2016-11-08      &  $J$         & 112~$\times$~30           & 1.04          &  0.6--1.1       & Clear                   \\
   2016-11-08      &  $K_s$        & 135~$\times$~10           & 1.15          &  0.6--1.1       & Clear            \\ \hline
\end{tabular}
\label{t:logGTC}
\begin{tabular}{l}
Standards: \\
OSIRIS: He3 ($F657/35$, $F802/51$, $F754/50$);  \\ OSIRIS: G191-B2B, Feige34, He3 ($F902/44$); \\
CIRCE:  AS05, AS16, AS39
\end{tabular}
\end{table}

\citet{1996ApJ...458..755R} reported detection of \psr in $\gamma$-rays with the {\it EGRET}
telescope aboard the Compton Gamma Ray Observatory.  
They found that the phase-averaged high-energy $\gamma$-ray pulsed emission from the pulsar 
has a single power-law  (PL) spectrum with the photon  index $\Gamma=2.8(3)$ and the integral flux 
$4.1(1.4)\times10^{-8}$ photons cm$^{-2}$ s$^{-1}$  above 100 MeV.
This was confirmed by the {\it Fermi} Large Area Telescope
(LAT) \citep{2013ApJS..208...17A}. The pulsar was detected with a pulsed flux of  
$7.1(6)\times10^{-8}$ photons cm$^{-2}$ s$^{-1}$ 
and a photon index of  $\Gamma=2.83(3)$ for the PL  and  $\Gamma=1.35(13)$ for the PLEC\footnote{Power law Super Exponential 
Cutoff model \citep{2013ApJS..208...17A}} spectral models, respectively, corresponding to the energy flux 2.7(1)$\times 10^{-11}$
ergs cm$^{-2}$ s$^{-1}$  in the 100 MeV to 100 GeV range \citep{2020ApJS..247...33A}.

The optical counterpart of PSR B0656+14 (V$\sim$25) was discovered by \citet{1994ApJ...422L..87C} 
and confirmed by \citet{1996ApJ...467..370P}, \citet{1997Ap&SS.252..451K} and
\citet{1997ApJ...487L.181S}.  
Later, \citet{1998A&A...333..547K} reported the first $BVRI$ photometric observations of the \psr
optical counterpart, which showed that 
the pulsar optical radiation has a non-thermal origin. The optical investigations were extended
using the \textit{HST} and the largest ground-based telescopes, 
such as the VLT. 
They afforded an opportunity to perform time-resolved photometry, to compile the detailed
UV-optical-IR spectral energy distribution 
and to measure the phase-averaged  and phase-resolved linear polarisation  
\citep{2000ApJ...543..318M, 2001A&A...370.1004K, 2003ApJ...597.1049K, 2005A&A...440..693S,
2006A&A...448..313S, 2007Ap&SS.308..545Z, 2011ApJ...743...38D, Mignani:2015aa}.

The total galactic interstellar colour excess $E(B-V)$ towards \psr obtained using the {\it COBE} and
{\it IRAS} data is 0.09 \citep{Schlegel:1998aa}. 
Based on analyses of hydrogen column densities from the radio and X-ray observations,
\citet{2006A&A...448..313S} concluded that the 
most likely value of the colour excess in the pulsar direction is $E(B-V)$ = 0.03. The new 3D
interstellar dust reddening model\footnote{http://argonaut.skymaps.info: it is based on analyses
of Pan-STARRS~1 and 2MASS stellar photometry} \citep{2015ApJ...810...25G, green2}, 
gives a total $E(B-V)$ of 0.06 and,
together with the pulsar parallax distance of 0.288 kpc, yields the colour excess towards \psr of
$E(B-V)$ = 0.01$^{+0.02}_{-0.01}$. 
Therefore, in the following analyses, we consider the interstellar absorption in the pulsar
direction $E(B-V)\le0.03$.
The low total galactic absorption and the small colour excess of the pulsar flux allow us to 
apply the dependence N$_{\mathrm H} = 8.3\times10^{21}$ cm$^{-2}$ $\times$ $E(B-V)$
\citep{2014ApJ...780...10L} to estimate the hydrogen column density towards the pulsar of
N$_{\mathrm H} \lesssim 2.5\times10^{20}$ cm$^{-2}$. 
Another estimation of hydrogen column density towards the pulsar is based on the empirical
relation between the hydrogen column density and optical extinction obtained by \citet{nh-ebv2016}
using the {\it Chandra} SNR archive. They found the empirical relation
N$_{\mathrm H}$  = 2.87$\pm$0.12$\times$ 10$^{21}A\mathrm{v}$ which gives 
N$_{\mathrm H} \lesssim$ 2.8$\times$10$^{20}$ cm$^{-2}$ to the pulsar using the 
standard reddening $A\mathrm{v}=3.2 E(B-V)$ law.  This N$_{\mathrm H}$ constraint shown by 
the dash-hatched  region in  Fig.\ref{FigX-rayFits}  can be used in spectral analyses of the
multi-wavelength data  from the near-IR through X-rays. First attempts   of such analysis for 
\psr were undertaken by \citet{2000ApJ...539..902E} and  \citet{PA-model2003} using a very 
limited  set of data available  at that time.   This  approach was shown to be  potentially
useful for obtaining    tighter  constraints  on   
the pulsar parameters as compared to those  inferred from separate spectral ranges. 
 Given a considerable amount of new near-IR-optical-UV and X-ray data obtained since that time,
it is useful to repeat  such analysis using a modern approach. 

\begin{figure*}
\setlength{\unitlength}{1mm}
\resizebox{15.cm}{!}{
\begin{picture}(130,145)(0,0)
\put (-5,73.){{\includegraphics[width=7.5cm, bb=-100 20 700 800,  clip=]{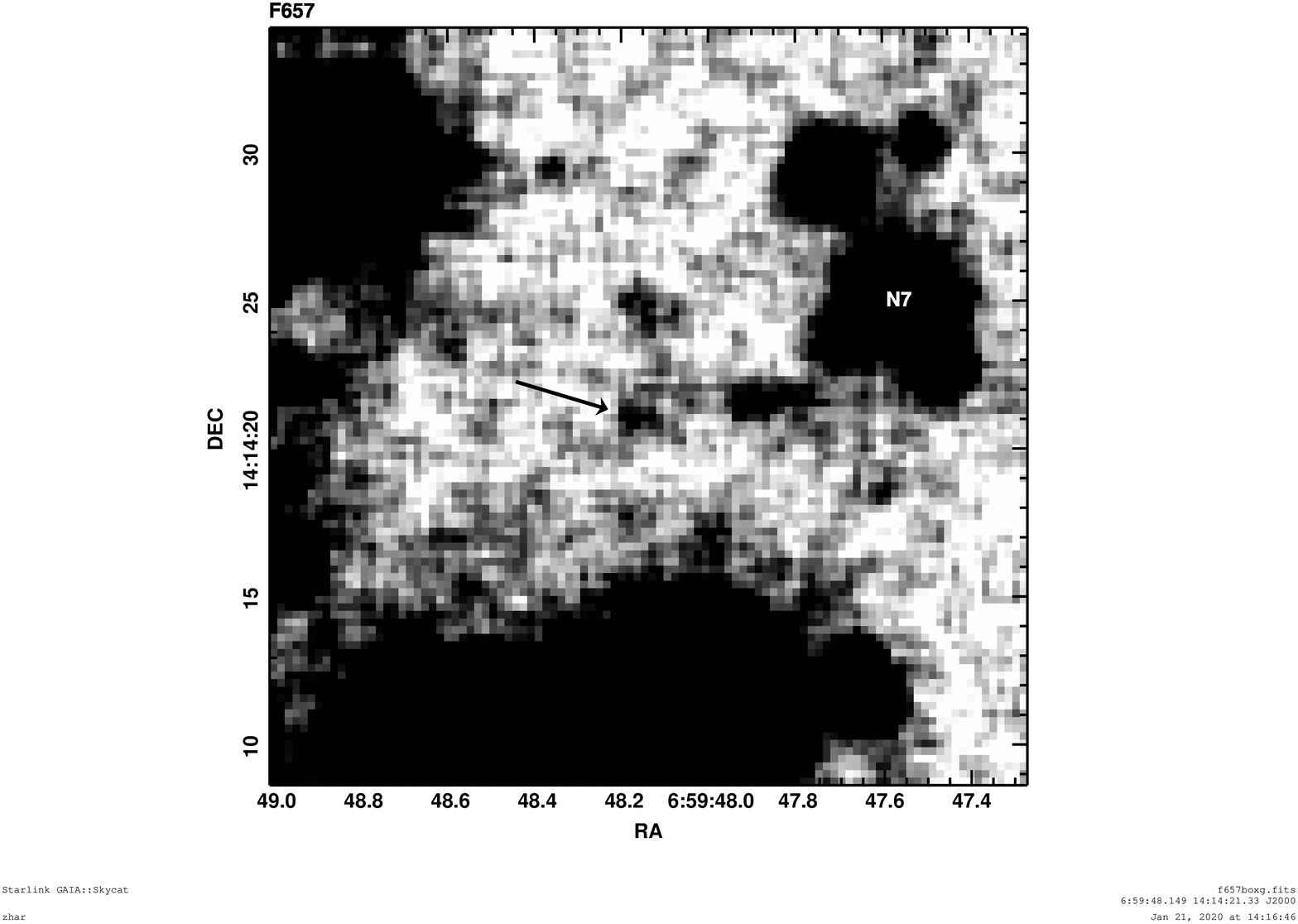}}} 
\put (69,73.){{\includegraphics[width=7.5cm, bb=-100 20 700 800, clip=]{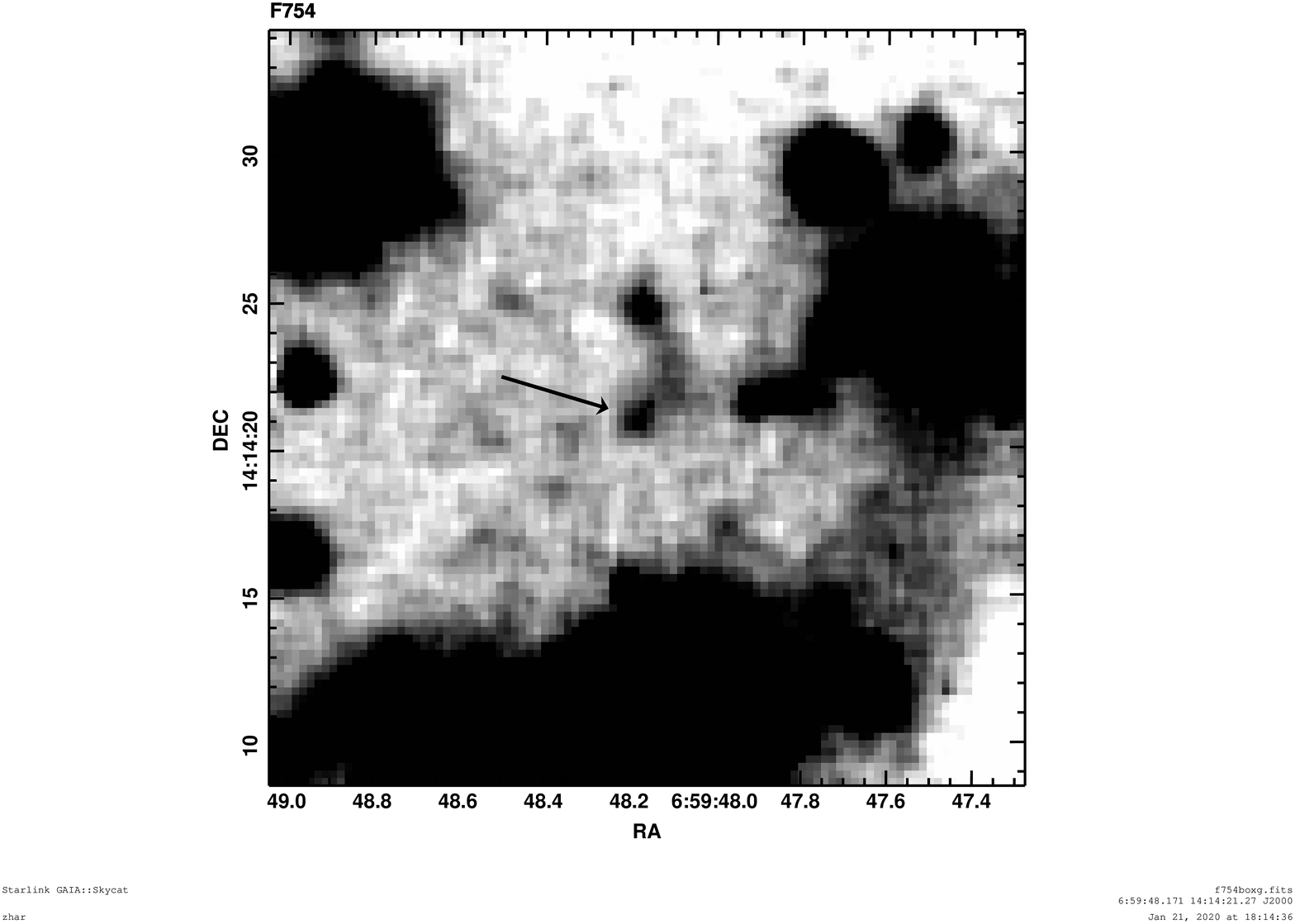}}}
\put (-5,0){{\includegraphics[width=7.5cm, bb=-100 20 700 800, clip=]{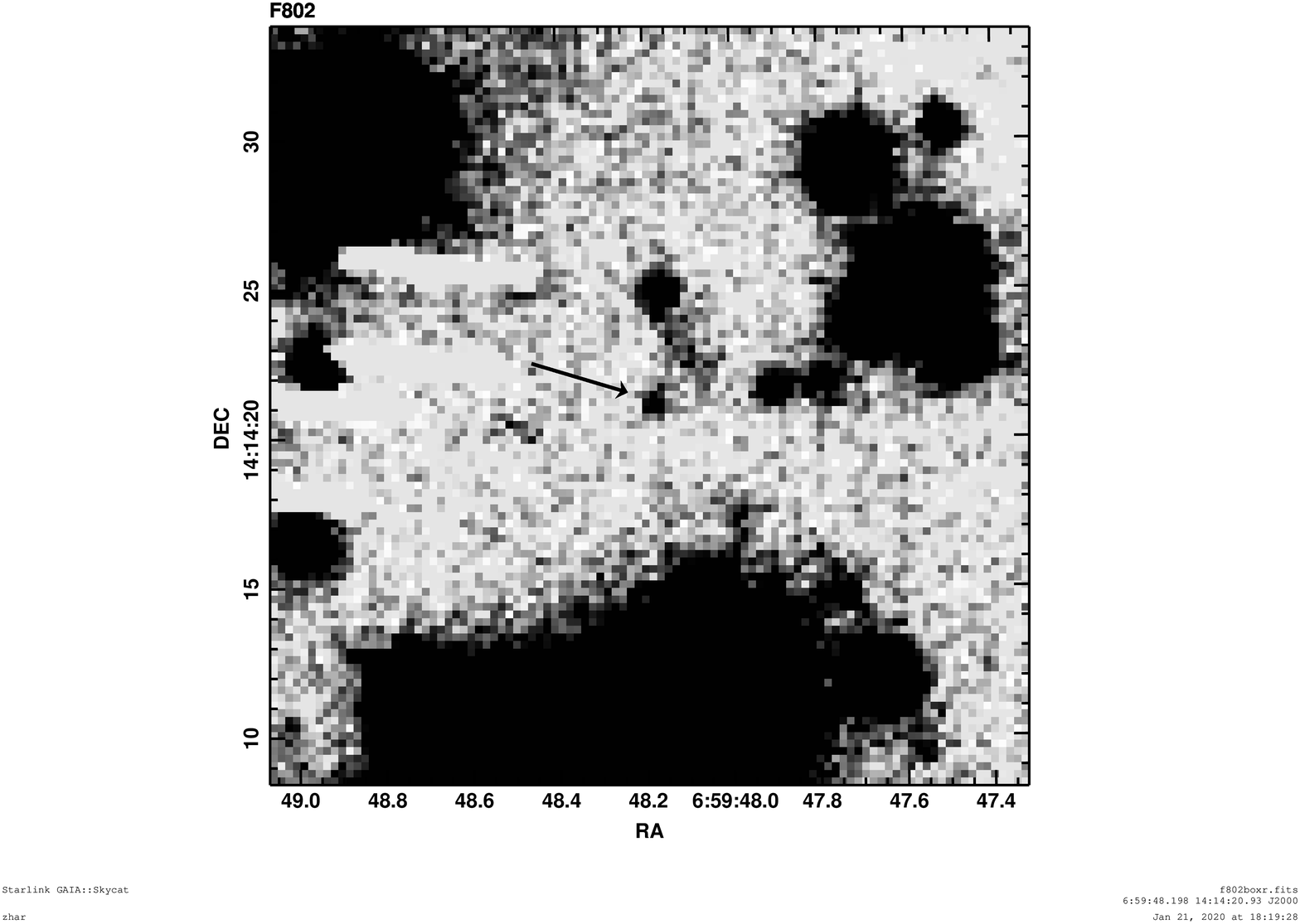}}} 
\put (69,0){{\includegraphics[width=7.5cm, bb=-100 20 700 800,clip=]{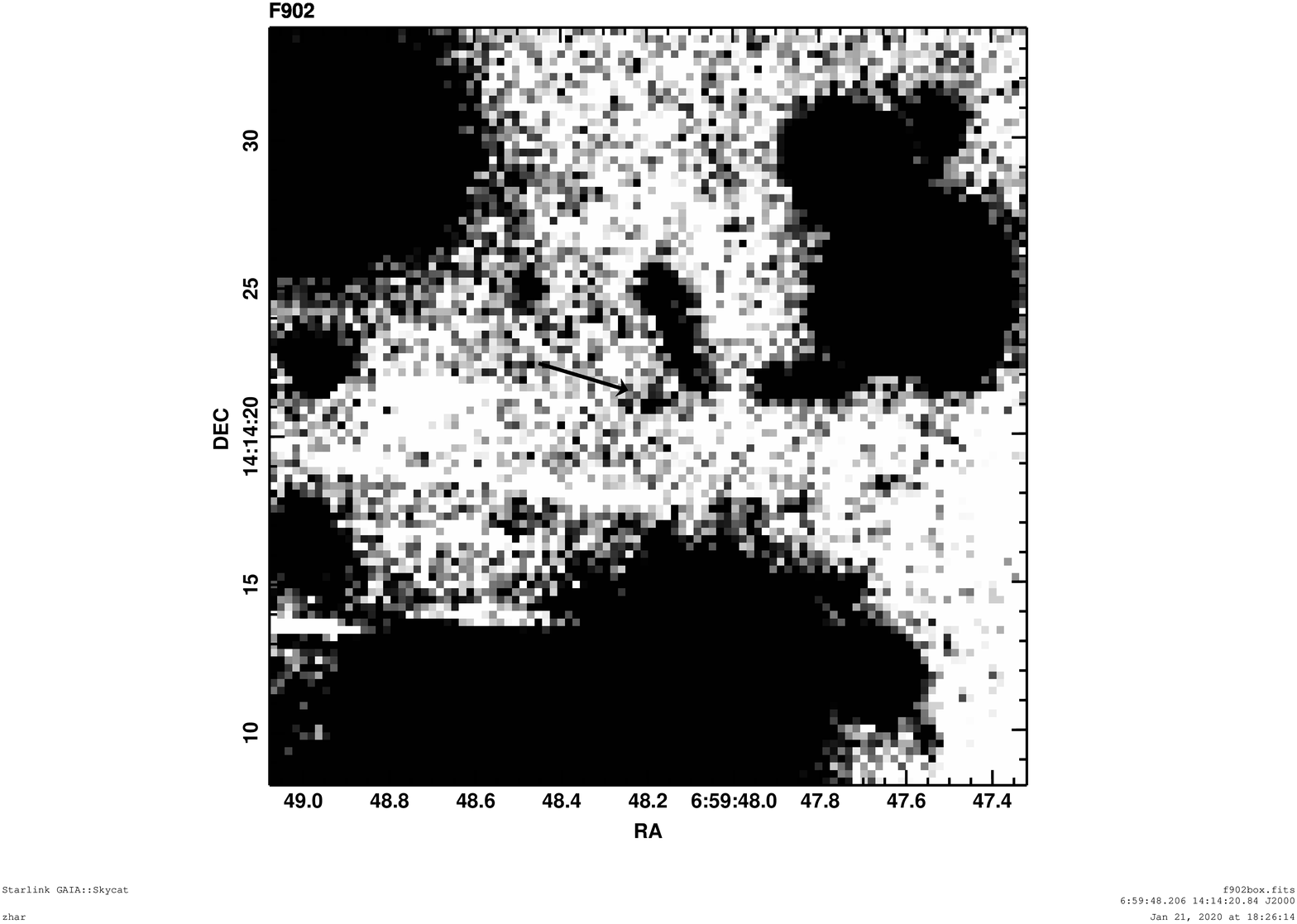}}}  
\end{picture}}
\caption{Fragments of the GTC/OSIRIS narrow-band images of the PSR B0656+14 field. 
The images are smoothed with a three pixel Gaussian kernel.
The pulsar optical counterpart is marked by the arrow. The star `N7' from  \citet{2001A&A...370.1004K} is marked too in the top-left panel (see discussion below).}
\label{fig:GTCn}
\end{figure*}

In this paper we   focus on such multi-wavelength data analysis for this pulsar. 
We include our
new optical narrow-band and near-IR $JHK$ band photometric observations 
obtained with the Gran Telescopio Canarias (GTC). We also present a detailed analysis of our VLT spectroscopic medium resolution observations of the pulsar 
whose preliminary results were only shortly reported in  \citet{2007Ap&SS.308..545Z}. For a consistent 
comparison with the VLT spectral 
and GTC photometric data  we   revised the pulsar narrow-band optical and broad-band near-IR data obtained 
with the HST and  presented by \citet{2011ApJ...743...38D}. We show that  hints of the optical absorption/emission  features in the spectrum of the pulsar reported by these authors are insignificant.  
We extracted  X-ray data from different missions  from archives,  including the recently obtained but 
not yet published \textit{XMM-Neuton} data, that double the number of the pulsar source counts in soft X-rays.  We re-reduced the data with modern versions of the respective tools, and 
 used them  to search for a common spectral solution across the IR-Optical-UV-X-ray domains.
The details of new GTC observations and data analysis are described in Section \ref{sec:obs}, 
 clarification
of the {\it HST} near-IR photometric results   using the GTC data is presented 
in Section~\ref{HST/NIC}, analyses of the VLT spectroscopy and reanalysis of the  HST optical
narrow-band photometric data are given in Sections~\ref{VLT/FORS2} and \ref{HST/ACS}. The X-ray
data  are described in  Section~\ref{sec:Xrays}.
The results and discussion are presented in Sections~\ref{sec:msf},\ref{sec:disc} and conclusions
in Section~\ref{sec:con}.

\begin{figure*}
\setlength{\unitlength}{1mm}
\resizebox{15.cm}{!}{
\begin{picture}(130,145)(0,0)
\put (0,-1){{\includegraphics[width=6.85cm,bb=-40 62 600 720,  clip=]{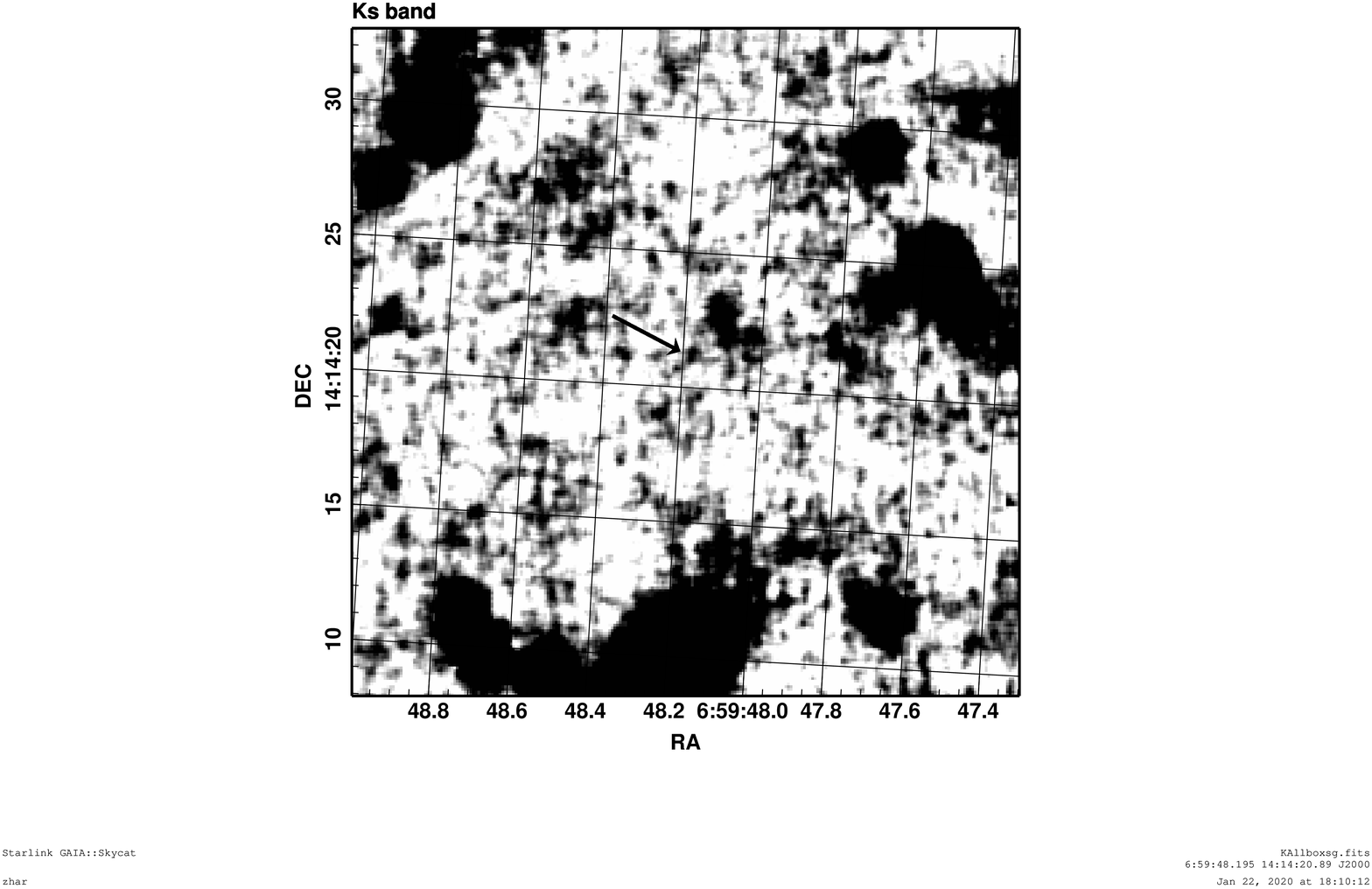}}}
\put (78,7){{\includegraphics[width=5.85cm, bb=155 245 695 800,  clip=]{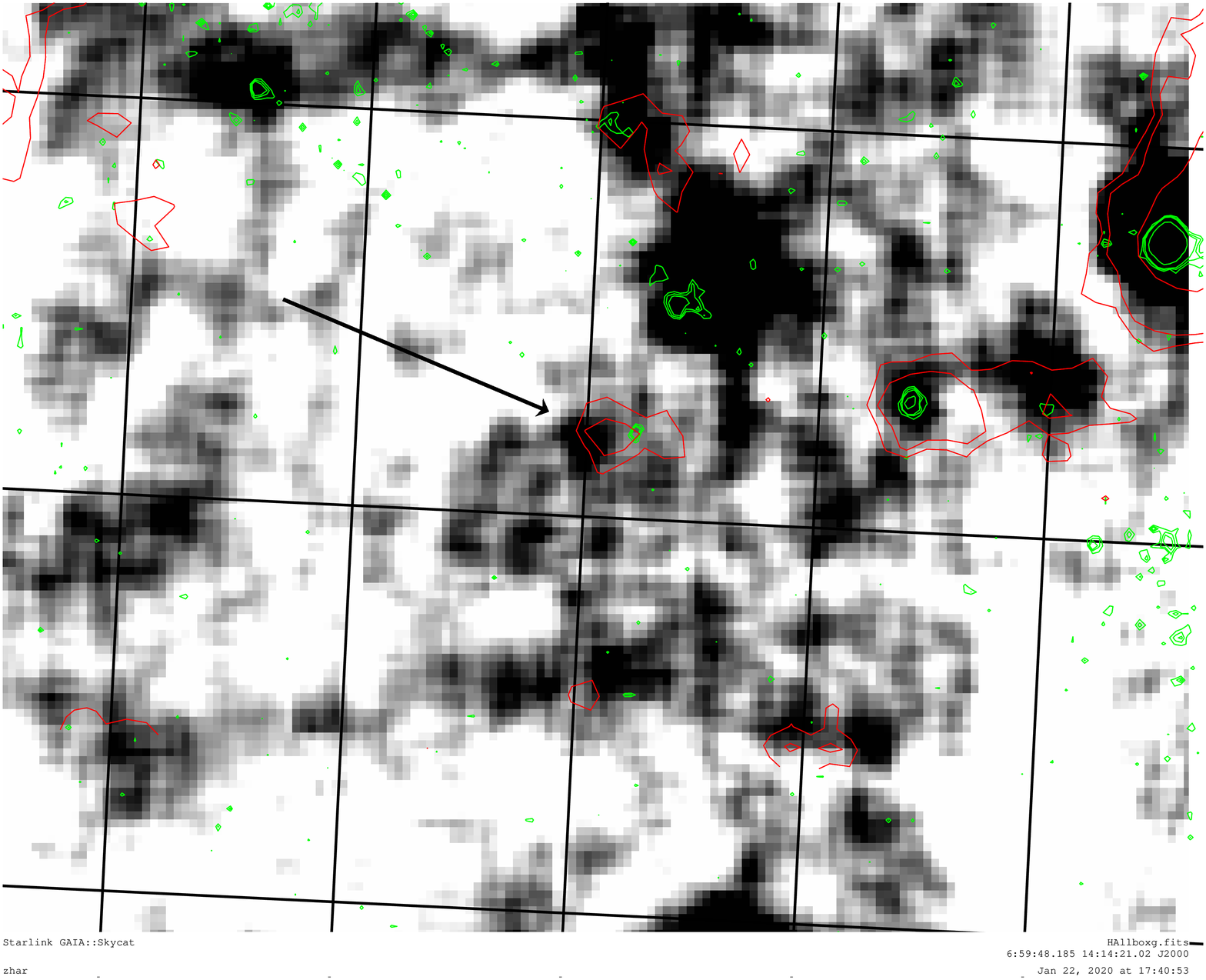}}}
\put (70,70.5){{\includegraphics[width=6.85cm, bb=-50 65 600 720,  clip=]{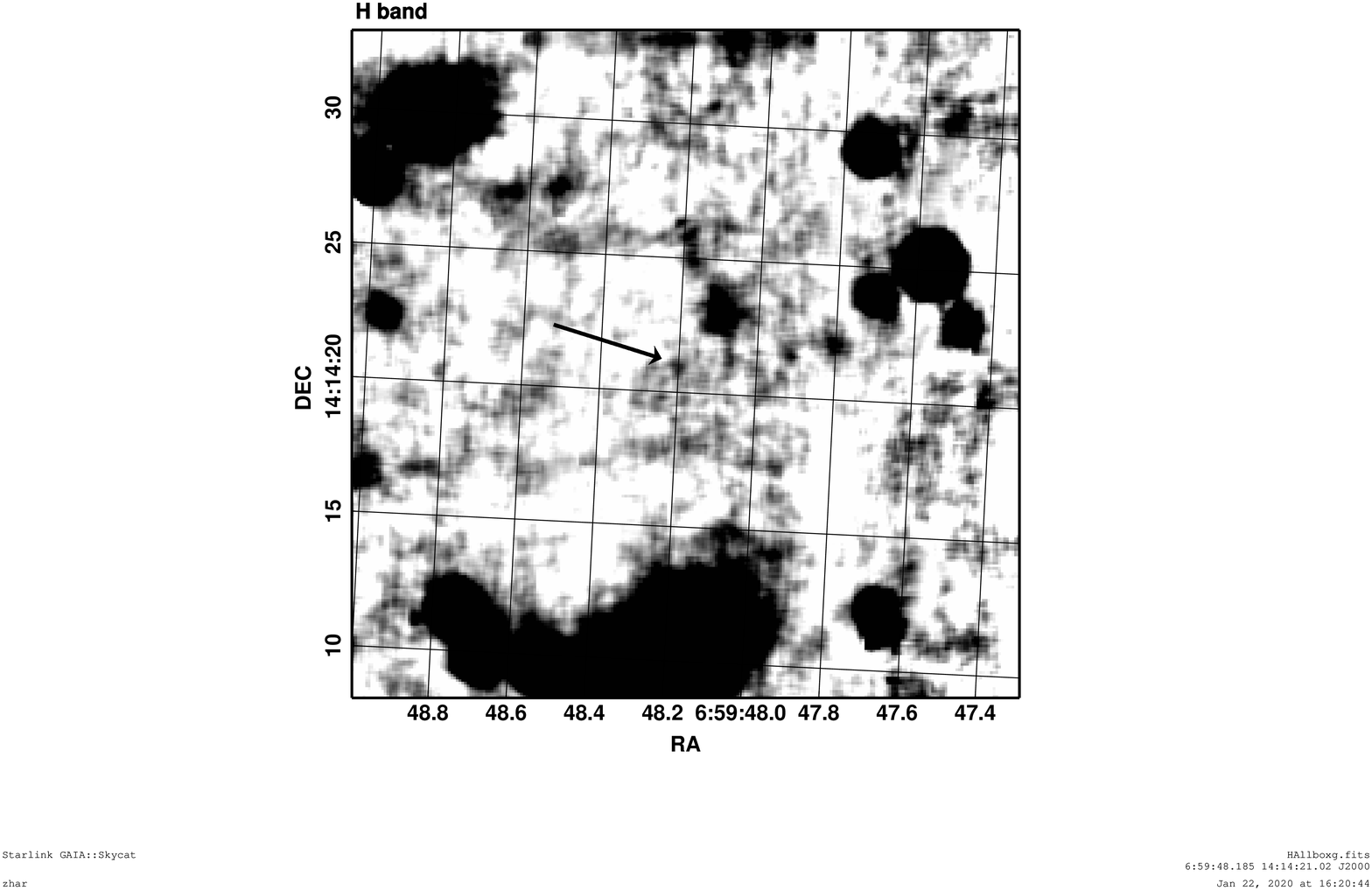}}}
\put (0,70.5){{\includegraphics[width=6.85cm,bb=-40 70 600 720,  clip=]{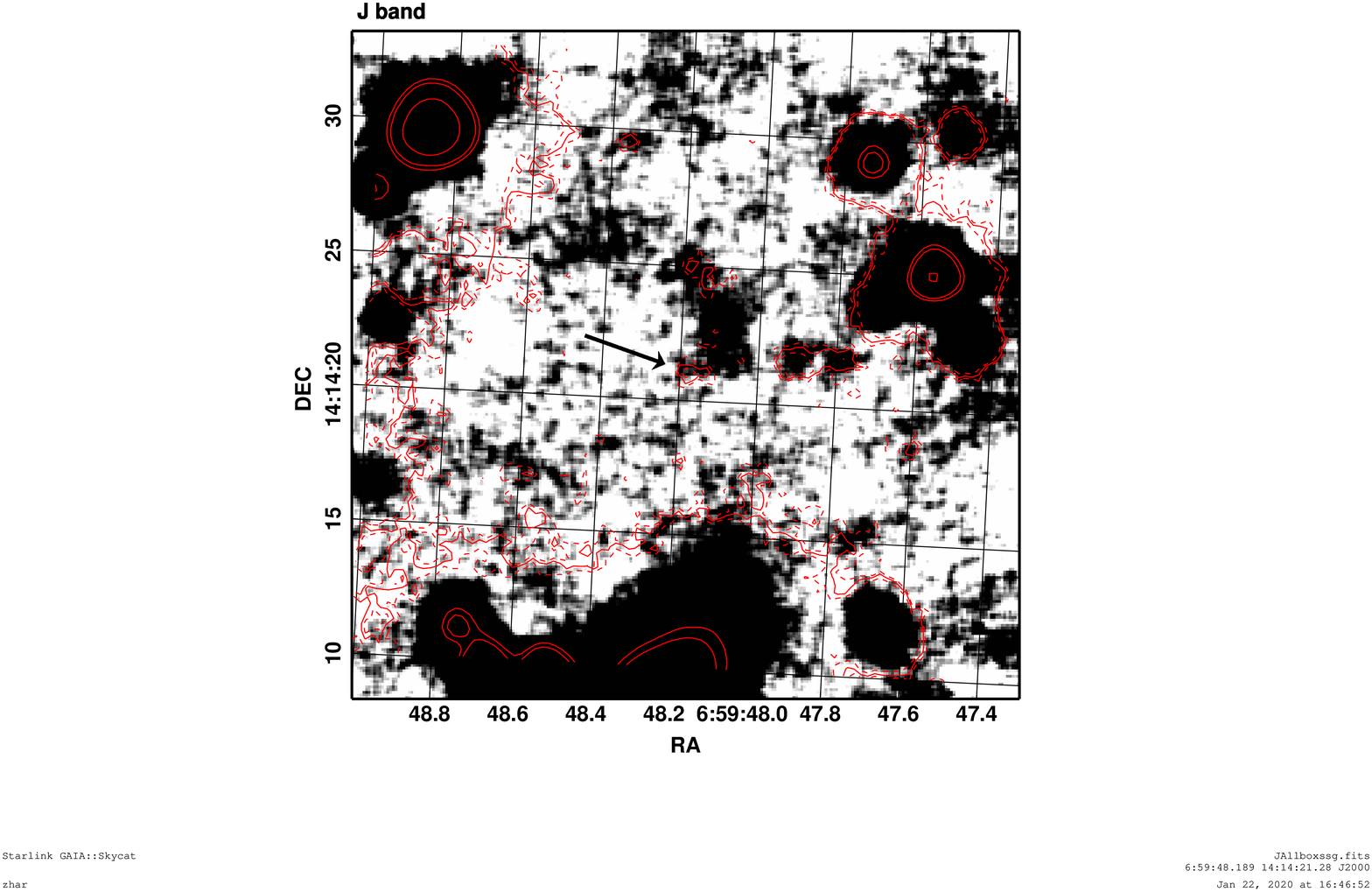}}}
\thicklines
\end{picture}}
\caption{Fragments of the GTC/CIRCE $JHK_s$ images of the PSR B0656+14 field.  The
$10^{\prime\prime}\times10^{\prime\prime}$ pulsar vicinity from the $H$-band image is enlarged in the right bottom
panel. All image fragments are smoothed with a three pixel Gaussian kernel.  The pulsar counterpart is marked  by
arrows.  The red contours are overlaid from the $F657/35$ band image (see Fig.\ref{fig:GTCn}) { (epoch 2011)}.
The green contours are overlaid from the {\it HST/FOC} $F110W$ image (see \citet{2001A&A...370.1004K,2011ApJ...743...38D}) { (epoch 1997)}.
The shift between the positions is due to the pulsar proper motion.}
\label{fig:GTCir}
\end{figure*}

\section{The GTC data}
\label{sec:obs}

\subsection{Optical narrow-band imaging}

The narrow-band photometric observations of \psr were obtained using the Optical System for
Imaging and low-intermediate Resolution Integrated Spectroscopy (OSIRIS\footnote{see
{http://www.gtc.iac.es/instruments/osiris/} for details}) in the $F657/35$, $F754/50$, $F802/51$ and $F902/44$ narrow
bands\footnote{The band pivot wavelengths and widths in {\sl nm} units are the first and  second numbers in 
filter ID names.} during several runs in 2009-2012\footnote{Proposals GTC4-09BCATMEX and GTC1-11BIACMEX,
PI~S.~Zharikov}. The instrument contains two CCDs with a plate scale of 0.254 arcsec/pixel (2$\times$2 binning)
and a total field of view (FoV) of 7\farcm8$\times$~7\farcm8. The log of observations is given in Table~\ref{t:logGTC}. 

Standard data reduction, including bias subtraction, flat-fielding and cosmic-ray removal, was
applied to the raw data using {\sc eso-midas} routines\footnote{https://www.eso.org/sci/software/esomidas//}. 
Some of the resulting images appeared to have detector defects in the pulsar location, and for
this reason were rejected. The total integration time for the resulting summed images was 2.1, 2.4, 3.6 and 
15.775 ks for the $F657/35$, $F754/50$, $F802/51$ and $F902/44$ bands, respectively. The images were astrometrically
calibrated using the {\sc iraf} {\it ccmap} task  and a set of eight reference stars 
in the vicinity of the pulsar from the {\it GAIA} catalog \citep{2016A&A...595A...1G,
2018A&A...616A...1G}. The {\it rms} error of the calibrations was  $\approx$0\farcs05 in both
directions. Selections of another sets of nearby  {\it GAIA} stars did not affect  this result
significantly. For the photometric calibration, spectrophotometric standard stars He3, G191-B2B and Feige34
\citep{oke74, stone77, massey, oke90} were observed each observing night in the corresponding bands (see
Table~\ref{t:logGTC}) and respective magnitude zero-points were obtained. 

The resulting images are presented in Fig.~\ref{fig:GTCn}. The pulsar optical counterpart is clearly detected on 
the combined images in all the four bands. Its instrumental magnitudes were measured in the aperture with a 
diameter of 4 CCD pixels (1\farcs0) centred at the pulsar position. They were then corrected for the PSF of bright
stars and transformed into the resulting pulsar magnitudes. The photometric results are presented in
Table~\ref{t:GTCphot} and Fig.~\ref{fig:PSRSpecRaw}.

\begin{table}
\caption{GTC photometry  of \psr.}
\begin{center}
\begin{tabular}{lcl}
\hline\hline
Band         & $\log (\nu)$            &.    Flux$_{obs}$                 \\
             &      [Hz]               &	 [{$\mu$Jy}]                 \\   \hline \hline                 
 $F657/35$   &   14.659        		   & 	 0.368(32)        	         \\
 $F754/50$   &   14.599        	       &  	 0.467(42)	                  \\ 
 $F802/51$   &   14.573      		   & 	 0.356(39)	                  \\
 $F902/44$   &   14.522      		   &  	 0.366(61) 	                 \\   
   $J$       &    14.393     		   &     0.469(130)             	  \\
   $H$       &   14.259      		   & 	 0.639(170)                  \\
   $K_s$     &    14.139      	       &	 1.059(300)                  \\  \hline
 \end{tabular}
\label{t:GTCphot}

\end{center}
\end{table}

\subsection{Near-IR broad-band imaging} 

The observations of the pulsar field were carried out in the $J$, $H$, and $K_s$ bands with the
Canarias InfraRed Camera Experiment (CIRCE, \citet{garner}) during several runs in 2016 October and 
November\footnote{Proposal GTC3-16BIACMEX, PI S. Zharikov} (see Table~\ref{t:logGTC}). CIRCE was operated as
a visitor instrument equipped with a 2048$\times$2048 engineering grade H2RG detector, which 
provides a resolution of 0.1 arcsec pixel$^{-1}$ and a FoV of 3\farcm4$\times$3\farcm4. 
To properly subtract the rapidly-varying near-IR sky background and 
reduce the effects of the poor detector cosmetic, we used a self-defined dithering pattern consisting of seven
positions. The individual exposures in the $H$ and $K_s$ 
bands were obtained using the detector Correlated Double Sampling (CDS) mode and were repeated two and three times at
each dither position, respectively. The $J$ band observations were performed in the detector Fowler sampling mode, and
the individual images were obtained twice at each dither position. The total integration time for the $J$, $H$, and $K_s$
bands was 9.24, 4.48 and 3.03 ks.  

The CIRCE data were reduced using custom codes written in {\sc idl}. All the images were dark subtracted (in particular to account for electronic offsets) and flatfielded, where the masterflat was produced as differential flat from a series of bright flats and faint flats. For sky-subtraction, for each image the adjacent image with the closer count-level was used.
After performing the primary reduction several bad pixel masks were applied to every individual exposure. This allowed us to 
fix numerous non-functional pixels, two dead amplifiers and other detector defects. 
Before combining the individual sky-subtracted images, they were aligned by measuring the centroid of one star in the FoV in all images. A simple mean combination was used to produce the final image. 
The resulting combined images are shown in Fig.~\ref{fig:GTCir}.
Their astrometric solutions  were obtained using a set of seven reference stars in the pulsar vicinity from the 
{\it GAIA} catalogue. The formal rms error of the astrometric calibration was $\approx 0\farcs02$ 
in both directions. A set of other stars  does not significantly  improve the  solution.

Inspection of the resulting   images  showed that the pulsar can be detected at a fair confidence level in the $H$
band and only marginally in the $J$ and $K_s$ bands. The latter is based on the position coincidence of the pulsar and
detected counts.  To demonstrate this we overlaid the image in the $F657/35$ band by contours on the $J$ band image
(Fig.~\ref{fig:GTCir}, top-left). The more reliable detection of the pulsar in the H band is explained by 
better seeing conditions during the observations in this band (see Table~\ref{t:logGTC}). A small $\sim$0\farcs2 dispersion
in the pulsar counterpart position in different near-IR bands is visible. However, it is a typical situation for objects
found near the detection limit (S/N$<$10) (see, for example, Figure 2 in \citet{2013MNRAS.435.2227Z}). 
We also note that the position of the pulsar in the H band perfectly agrees with the expected pulsar radio position 
calculated for the epoch of the near-IR observations using its  proper motion measured in the radio by
\citet{Brisken:2003aa}. Contours in the bottom-right panel of Fig.~\ref{fig:GTCir} demonstrate the proper motion shift
with the epoch observed in the optical-near-IR.   

The conditions during the GTC/CIRCE observations were clear with slightly ($\sim$0.1 mag) variable transparency. 
The photometric calibration was performed based on the standards from the photometric standard
fields AS05,  AS16 and AS39 from \citet{hunt}  observed the same nights as the target. In addition, we used the
Two-Micron All-Sky Survey ({\it 2MASS}) stars \citep{2mass} with a magnitude of $\la$~14.5 and photometric errors
$\la$~0.05 mag that fall in the CIRCE FoV. We accounted for the atmospheric extinction using the coefficients 
0.09, 0.07 and 0.08 for the $J$, $H$, and $K_s$ bands, respectively, provided by the GTC team. As a result, we obtained the 
photometric zero-points $J^{ZP}=23.74\pm{0.03}$, $H^{ZP}=24.20\pm{0.04}$ and $K_s^{ZP}=23.92\pm{0.04}$. 

We measured the pulsar instrumental magnitudes in the  
aperture with a diameter of 6 CCD pixels (0\farcs6) centred at the pulsar position
computed for the epoch of the near-IR observations. They were corrected for the PSF of bright stars selected in 
the CIRCE FoV and transformed into the {\it 2MASS} $JHK_s$ magnitudes using the obtained zero-points. 
The magnitudes were then transformed into fluxes using the calibrations from \cite{Cohen:2003aa}.
The photometric results are presented in Table~\ref{t:GTCphot} and Fig.~\ref{fig:PSRSpecRaw}. 

\begin{figure}
\includegraphics[width=8.5cm, clip=]{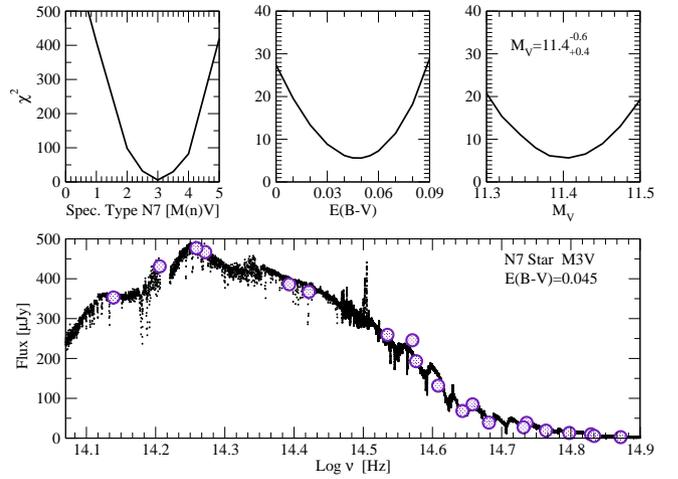}
\caption{The multi-band {\it HST/ACS/WFC}, BTA, {\it HST/NICMOS} and {\it 2MASS} photometry (filled circles) 
of the reference star N7 and its comparison with a template spectrum (points) of the M3V star (from
\citet{Rayner:2009aa} and  \citet{Kesseli:2017aa}).}
\label{fig:N7Spec}
\end{figure}

\section{Clarification 
of the HST/NICMOS near-IR photometry} 
\label{HST/NIC}

The {\it HST/NICMOS} photometry of \psr in the $F110W$, $F160W$ and $F187W$ bands was first reported by
\cite{2001A&A...370.1004K} with the respective fluxes 0.336(30), 0.575(30) and 0.779(97) $\mu$Jy.
\citet{2011ApJ...743...38D} have performed independent measurements based on the same data and obtained the 
fluxes $F110W$=0.385(30) $\mu$Jy, $F160W$=0.551(30) $\mu$Jy and $F187W$ =0.660(25) $\mu$Jy.
We verified these results using the original data from the {\it HST} archive and PHOTFNU parameters
extracted from the file headers and obtained from the STScI  
webpage\footnote{www.stsci.edu/hst/nicmos/perfomance/photometry/prencs\_keywords.html}.
Our measurement of $F160W$=0.563(30) $\mu$Jy is in a good consistency with the respective $F160W$-band
flux reported by \cite{2001A&A...370.1004K}. The flux in the $F110W$ band equals to 0.340(30) $\mu$Jy for
PHOTFNU=1.84724e-6 (CAMERA2) from the NICMOS webpage and  
0.393(30) $\mu$Jy for PHOTFNU=2.3133694e-6 from the image header.  
The first value corresponds to the measurement of \citet{2001A&A...370.1004K} and the second one
corresponds to that of \citet{2011ApJ...743...38D}.
We used a sum of all images to estimate the flux in the $F187W$ band. It is important to note that a
few cosmic rays were detected close the pulsar position in some images and they significantly complicated the
pulsar flux measurements. We verified each image for the presence of  
cosmic rays at or close to the pulsar position and removed them by interpolating the average
background pixel flux in the pulsar vicinity. 
Our resulting estimation of the pulsar flux in this band, $F187W$ =0.668(120) $\mu$Jy, is in
agreement with the \citet{2011ApJ...743...38D}   
measurements, however we note that they underestimated the flux errors. A slight shift (1 pixel)
of the photometric aperture centre  leads to a significant ($\sim$\%20) variation of the measured flux. 

As was shown above, the mentioned NICMOS calibration 
uncertainty can lead to the pulsar flux uncertainty above $2\sigma$ of statistical errors. 
To verify the calibration, it is instructive to consider a bright {\it 2MASS} reference star located
near the pulsar. To do that, \citet{2011ApJ...743...38D}  remeasured the fluxes of the object N7
(2MASS J06594760+1414253, see Fig.~\ref{fig:GTCn}) from \citet{2001A&A...370.1004K} 
and found them to be about 10\% higher than those reported by \citet{2001A&A...370.1004K} and
about 20\% lower than the {\it 2MASS} measurements ($J_{2MASS}$=16.37(14), 
$H_{2MASS}$=15.56(16), $K_{2MASS}$=15.13(16)). In our CIRCE data this object was detected
with a high S/N ratio and its magnitudes and colour indices in the {\it 2MASS} system are $J$=16.54(4), $H$=15.83(4),
$K$=15.62(4), $J-H$ = 0.72(5) and $H-K$ = 0.20(4). The high precision optical photometry of the object N7 was presented
by \citet{1998A&A...333..547K} and \citet{2001A&A...370.1004K}: $B$=22.00(3), 
$V$=20.31(2), $R_c$=19.13(2), $I_c$=17.74(2), $B-V$ = 1.69(4), $V-R$ = 1.18(3), $R-I$ = 1.38(3)).
The corresponding total Galactic extinction values in the \psr direction are $A_B$=0.32,
$A_V$=0.24, $A_R$ = 0.19, $A_I$=0.14, $A_J$=0.06, $A_H$=0.04 and $A_K$=0.03 \citep{Schlafly:2011aa}.
We repeated the {\it HST/NICMOS} photometry of the N7 object using the individual frames where the
object does not fall on the detector edge. The {\it PHOTFNU} parameters were extracted from the image headers. 
The measured object fluxes are $F110W^{N7}$=320.20(30) $\mu$Jy, $F160W^{N7}$=451.32(30) $\mu$Jy and 
$F187W^{N7}$ =351(2) $\mu$Jy.

Taking into account the object magnitudes, colour indices and the total Galactic extinction, we
found that the object N7 is likely a M0 --- M5 type dwarf star located at a distance of about 
750$\pm$300 pc \citep{Hawley:2002aa, Koen:2002aa}. The {\it GAIA}
parallax\footnote{https://www.cosmos.esa.int/web/gaia/data-access} of the N7 star is
1.88(38) mas which corresponds to a distance of $531^{+135}_{-89}$ pc.
The corresponding interstellar  extinction for this distance from 
 \citet{2015ApJ...810...25G, green2}  yields a colour excess of $E(B-V)$ = 0.02(2).
To check what fluxes can be expected from the star in the HST/NICMOS 
filters, we calculated them using the M0V-M5V type star IR spectral templates from
\citet{Rayner:2009aa}. They were extended to the optical range using data from the 
empirical template library of the Sloan Digital Sky Survey stellar spectra \citep{Kesseli:2017aa}
using an overlapping region. The best agreement between the observed and calculated colour indices
was  found for the M3V type star spectra ($M_V=11.4^{-0.6}_{+0.4}$, $E(B-V)=0.045(20)$), which are
consistent within uncertainties  
with our GTC/CIRCE photometry of the N7 star (see Fig.~\ref{fig:N7Spec}). Using only $JHK_s$ data
gives reduced  chi-square per degree of freedom $\chi_{\nu=3}^2\approx1$. The optical BTA 
photometry data (see Section~\ref{BTAdata}), after being included in the fit, insignificantly increases the reduced chi-square 
$\chi_{\nu=7}^2= 5.6$ that is more probably related to band-pass characteristics in the optical range
or the star individuality. 
The narrow band {\it HST/ACS/WFC} photometry of the N7 star is also in agreement with the N7 star
spectral type (the {\it  HST/ACS/WFC} data of the N7 star are presented in Fig.~\ref{fig:N7Spec}).

 The best fit  shows a full consistency ($\sim$1\%) between the measured and calculated fluxes in the
 $F160W$ band and a slight discrepancy of $\approx$10\% and $\approx$12\% for the respective 
 measurements\footnote{The magnitudes were obtained by convolution of the NICMOS
filters transmission curves with the template spectrum and NICMOS zero-points.} in the $F110W$ and $F187W$ bands:
$F110W^{N7}_{calc}$=353.0 $\mu$Jy, $F160W^{N7}_{calc}$=447.0 $\mu$Jy and $F187W^{N7}_{calc}$
=394.0 $\mu$Jy. We assume that the flux difference should be taken into account and that the NICMOS fluxes
obtained using {\it PHOTFNU}~ from the image header must be corrected accordingly.
After applying this correction   we obtain the observed pulsar fluxes 
$F110W$=0.432(30) $\mu$Jy, $F160W$=0.568(30) $\mu$Jy and $F187W$ =0.748(120) $\mu$Jy (Fig.~\ref{fig:PSRSpecRaw}). 

\begin{table}
\begin{center}
\caption{Log of the VLT/FORS2 observations of PSR B0656+14.}
\begin{tabular}{lcccl}
\hline\hline
 Date              & Number      &  AM         &   Exposure    \\
    (UT)            & of exposures                &                  & [s]   \\
   \hline \hline                 
   2004-11-12      &  2     &  1.30         &     1400             \\
   2004-11-14      &  4     &  1.30         &     1400             \\  \hline
   2004-11-15      &  2     &  1.30         &     1400              \\
   2004-11-16      &  2     &  1.29         &     1400              \\   
   2004-11-17      &   2    &  1.29         &     1400              \\
   2004-12-17      &   4    &  1.36         &     1400              \\
   2005-02-07      &    2   &  1.33         &     1400               \\ \hline
\end{tabular}
\label{t:logVLT}
\end{center}
\end{table}

 \begin{figure}
 \setlength{\unitlength}{1mm}
\resizebox{15.cm}{!}{
\begin{picture}(130,96)(0,0)
\put (0,37){\includegraphics[width=7.6cm, clip=]{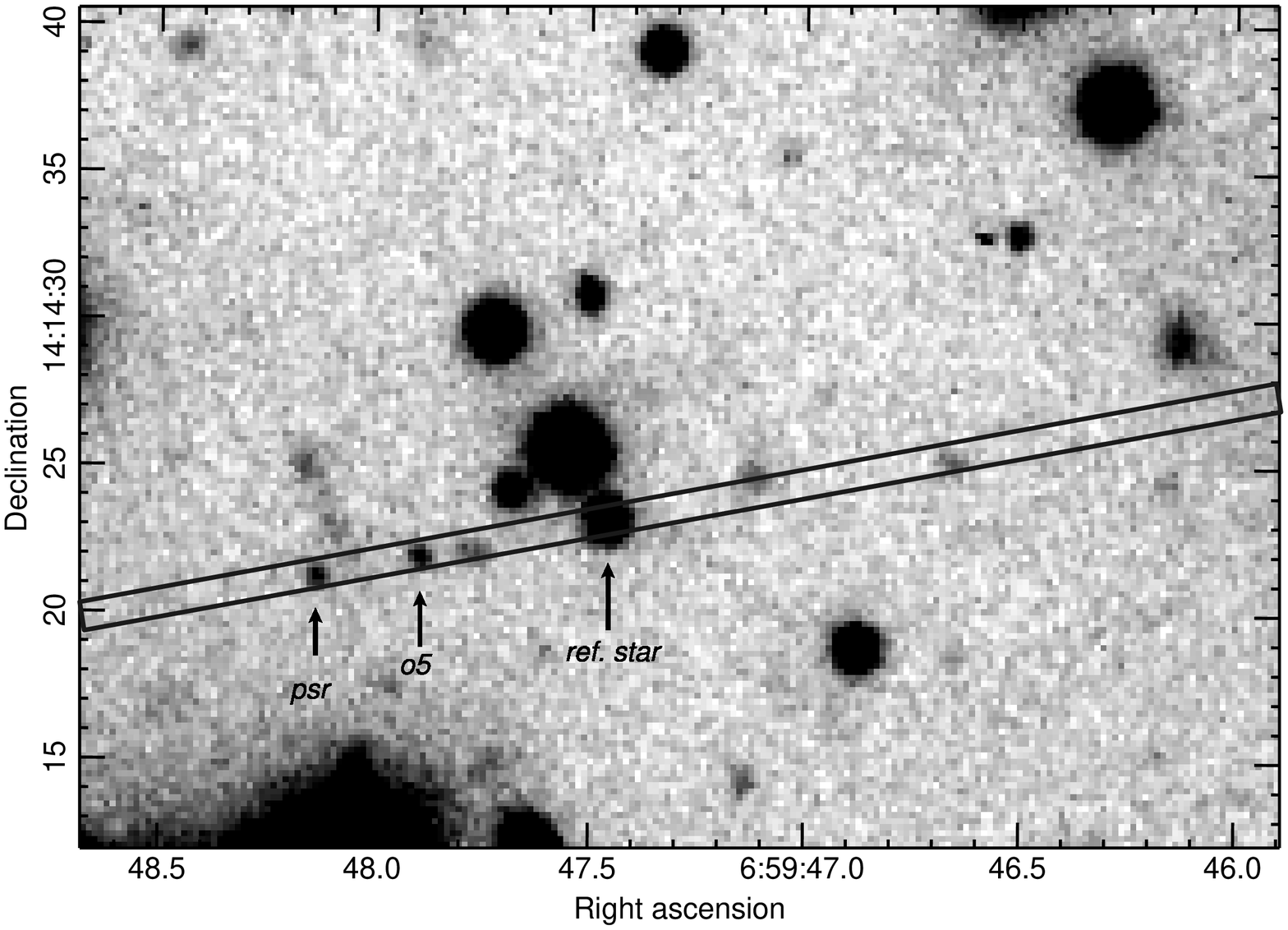}}
\put (6,-1){\includegraphics[width=6.9cm, height=4cm, bb= 45 180 540 580,  clip=]{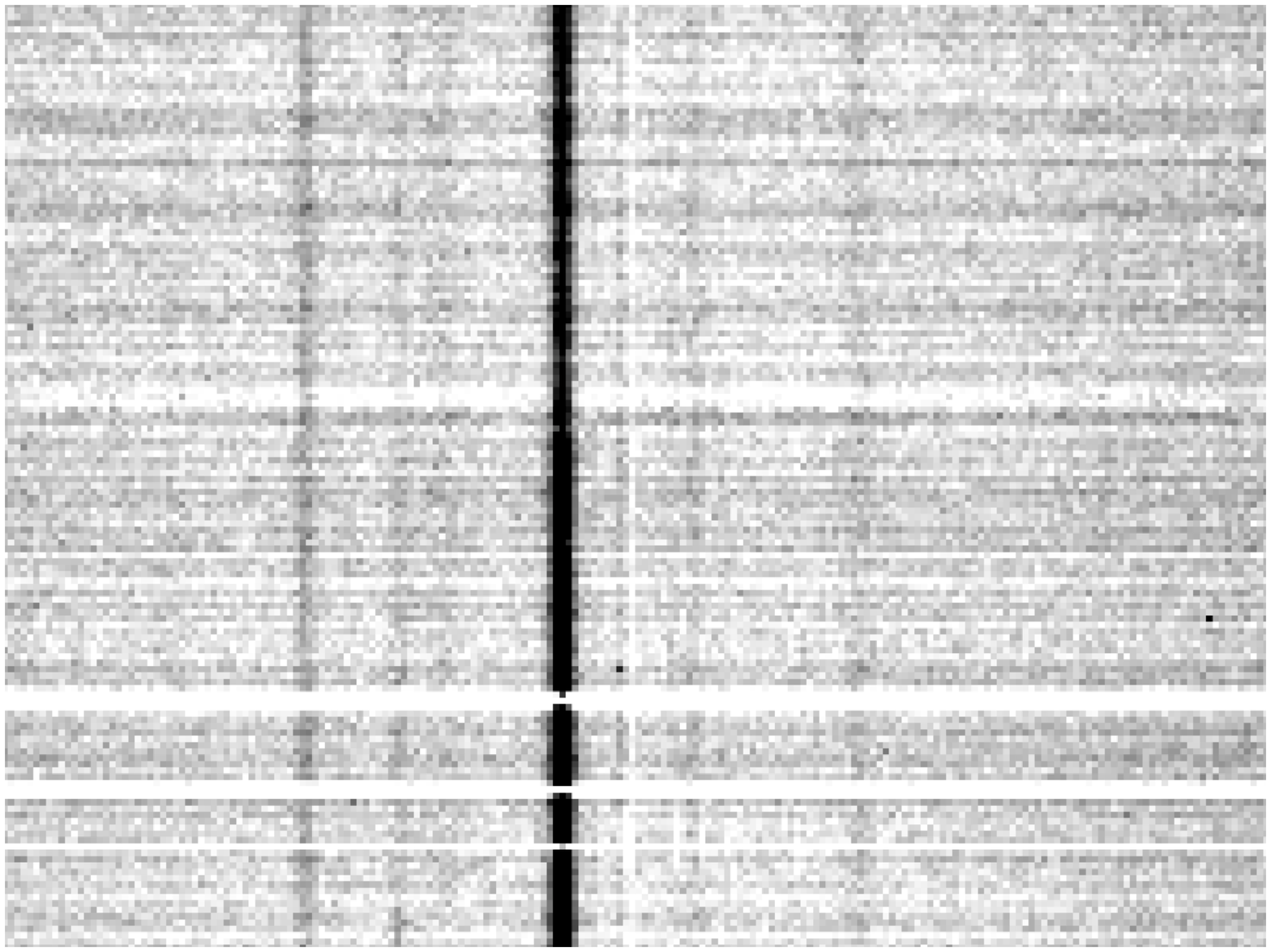}}
\end{picture}}
\caption{{\sl Top:} $R$-band image fragment of the pulsar field. The stripe shows the slit and its
orientation, which were used to obtain spectra of the pulsar and two nearby stars marked as `psr', `o5' and `ref. star'.
{\sl Bottom:}  fragment of the VLT/FORS2 2D-spectrum of the pulsar and nearby objects in the `blue' optical range.  }
\label{fig:GTCslit}
\end{figure}

\begin{table*}
\caption{Log of the {\it HST/ACS/WFC} observations of PSR B0656+14 and its flux measurements in the Ramp
filter sub-bands.}
\begin{tabular}{lcclllllllcccc}
\hline\hline
   Exp.       &    $\log (\nu_{C.W.})$  &   {\scriptsize PHOTFLAM}                 & Flux    & Flux       &Flux     & Flux(2)    & Flux(3)      & Flux(4)   & Flux$_{ave}$ & {\it Ap.}                 & {\it CTE }          & Flux$_{obs}$  & Flux$_{obs}$   \\
   Time      &                                     &   (2009)                                      &$2pix$                 &$3pix$                  &$4pix$                 &  $ 0$\farcs5         &  $0$\farcs5           & $0$\farcs5          & $0$\farcs5          & {\it corr.} $\infty$  &   {\it  corr.}      &      obs                  &     derredded                   \\                  
  (sec)    &    (Hz)                          &    $(\frac{ergs}{cm^2 \AA e^-})$ & (cps)  &  (cps) & (cps) & (cps)  &  (cps)  & (cps)  & (cps)  & \%                   & \%             &       ($\mu$Jy)      &     ($\mu$Jy)          \\ \hline \hline                                                                                                                                                                        
	1399.8$^\circ$ & 	14.872(11)$^*$              &   1.743e-18   &0.191&  0.255&   0.316&  0.282&  0.307& 0.348 &0.312(26) &11.7  &5.6  &0.347 & 0.387(34) \\               
	1020.0$^\circ$ & 	14.834(11)$^*$	             &   9.520e-19   &0.293&  0.475&   0.629&  0.427&  0.569& 0.69&0.56(10) & 9.6   &3.6     &0.393 & 0.433(83) \\		  
	830.0$^\circ$  & 	14.798(12)$^*$	             &   7.162e-19   &0.410&  0.472&   0.542&  0.594&  0.565& 0.596&0.585(14)  & 9.1   &4.0  &0.362 & 0.396(10)\\	            
	885.0$^\circ$   & 	14.764(13)$^*$	             &   5.776e-19   &0.412&  0.461&   0.481&  0.591&  0.550& 0.532 &0.557(24)&8.7   &3.7  &0.323 & 0.350(15) \\		   
	1020.0$^\circ$ & 	14.732(12)$^\star$	    &   5.857e-19   &0.438&  0.483&   0.608&  0.599&  0.568& 0.664 &0.610(40) & 8.4   &3.0   &0.413 & 0.443(29) \\		   
	1020.0$^\circ$ & 	14.682(11)$^\star$	    &   3.779e-19   &0.500&  0.586&   0.610&  0.701&  0.694& 0.672  & 0.688(11) &8.5   &2.5  &0.376 & 0.402(6) \\            
	810.0$^\bullet$   & 	14.643(12)$^\star$       &   2.895e-19   &0.635&  0.732&   0.684&  0.891&  0.864& 0.751 & 0.835(61) &8.3   &2.5   &0.417 & 0.442(32)\\		   
	810.0$^\bullet$   & 	14.608(13)$^\star$       &   2.551e-19   &0.499&  0.700&   0.819&  0.790&  0.884& 0.916 &0.863(53) &8.5   &2.5    &0.447 & 0.470(29) \\		   
	747.0$^\bullet$   & 	14.576(14)$^\dagger$  &   2.427e-19   &0.466&  0.620&   0.697&  0.761&  0.788& 0.785 & 0.777(11) &8.8   &3.2   &0.448 & 0.468(6)\\		   
	1330.8$^\bullet$ & 	14.535(14)$^\dagger$  &   2.970e-19   &0.325&  0.366&   0.402&  0.623&  0.500& 0.476 & 0.533(64) &9.7   &2.9    &0.457 & 0.474(57) \\		   
	2842.0$^\bullet$ & 	14.500(17)$^\dagger$  &   5.474e-19   &0.089&  0.098&   0.133&  0.204&  0.153& 0.174 &0.177(21) &19.8 &3.2    &0.360 & 0.372(43) \\ \hline	   
\end{tabular}
\begin{flushleft}
\begin{tabular}{l}
E(B-V) = 0.03 is selected for flux dereddening;
observational epochs:  $^\circ$2005-12-08; 
$^\bullet$2005-12-03;  
Ramp filter names: $^*$FR459M;
$^\star$FR647M; 
$^\dagger$FR914M \\
$\log \nu_{CW}$ - logarithm  of frequency corresponding to the filter central wavelength. \\
Flux(N) is
the flux obtained from N pix. aperture by correction to 0\farcs5 aperture. \\
Flux$_{ave}$ = $\frac{\Sigma Flux(N)}{3}$ \\
{\it Ap. corr.} is the  correction for the PSF from the 0\farcs5 aperture to the nominal `infinite' aperture of 5\farcs5. \\
{\it CFE} is the charge transfer efficiency. \\
The final errors include errors of measurements and  uncertainties of the interstellar absorption. 
\end{tabular}
\end{flushleft}
\label{t:logHST}
\end{table*}

\begin{figure}
\setlength{\unitlength}{1mm}
\resizebox{15.cm}{!}{
\begin{picture}(130,38)(0,0)
\put(-1,0) {\includegraphics[width=3.4cm, bb = 0 0 650 700, clip=]{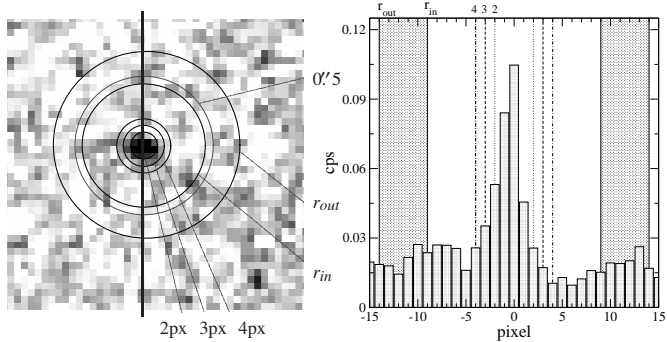}}
\put(35,-2) {\includegraphics[width=4.1cm, bb = 20 20 550  550, clip=]{zharFig6b.eps}}
\put (16.5,0.5){\scriptsize 2px}
\put (21,0.5){\scriptsize 3px}
\put (25.5,0.5){\scriptsize 4px}
\put (34.,7){\scriptsize $r_{in}$}
\put (34.,15){\scriptsize $r_{out}$}
\put (34.,29){\scriptsize 0\farcs5}
\put (14.5,2.5){\line(0,1){35}}
\end{picture}}
\caption{Example of HST/ACS/WFC image (left) of the pulsar obtained in the filter FR647M with the central wavelength 
(CW) 
corresponding to the frequency 
${\log \nu_{C.W.}=14.732}$~Hz and the pulsar spatial profile (right). The circular  apertures,   background region, and the 0\farcs5
aperture selected for the pulsar photometry are shown in the image. They are also shown by the vertical lines and grey strips in the
profile.} 
\label{fig:ACS}
\end{figure}

\begin{figure*}
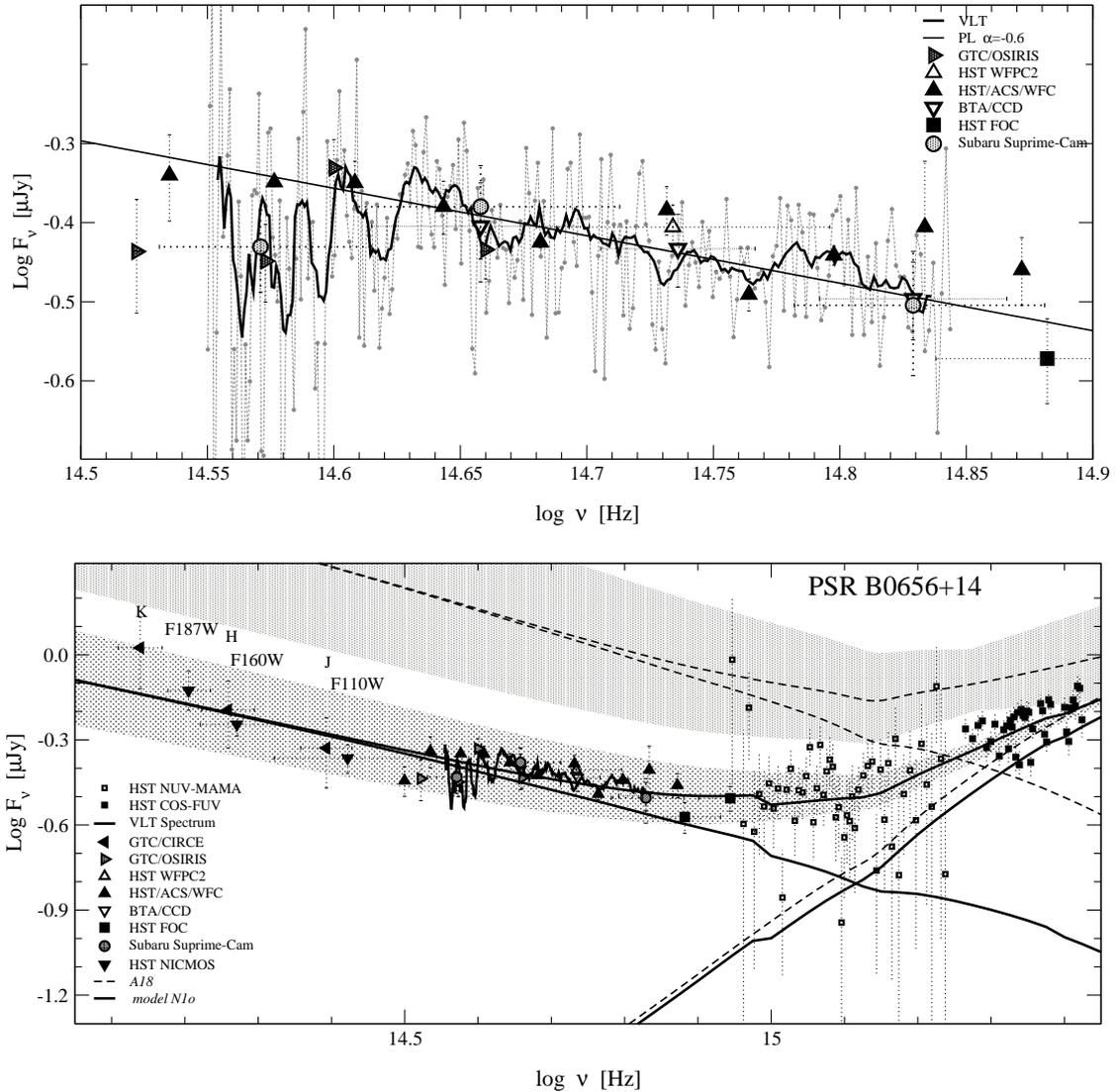

\setlength{\unitlength}{1mm}
\resizebox{15.cm}{!}{
\begin{picture}(130,127)(0,0)
\put(1.5,65.5) {\includegraphics[width=12.75cm, clip=]{zharFig7a.eps}}
\put(0,0) { \includegraphics[width=12.75cm, clip=]{zharFig7b.eps}}
\end{picture}}
\caption{{\sl Top:} observed VLT optical spectrum of \psr shown by the line and broadband photometric points (error-bars) 
obtained with different telescopes/instruments as notified in the legend. The straight line shows the best fit of the 
spectrum by the PL model with 
the spectral index $\alpha \approx 0.6$. {\sl Bottom:} observed UV-optical-near-IR spectrum of the pulsar produced compiling 
the data from various telescopes discussed in the text.  Low-wavelength extrapolations  of the absorbed thermal and 
nonthermal spectral components and their sum obtained from fitting the pulsar X-ray data  alone by the G2BB+PL model with
the absorption line (Table~\ref{tab:Fit:M1}, the fit \textit{N1}). 
Solid lines show the same when the long-wavelength and X-ray data are fitted together (the fit \textit{N1o}). Grey regions 
show 1$\sigma$ uncertainties of the fits  (see text for details). 
}
\label{fig:PSRSpecRaw}
\end{figure*}

\section{VLT/FORS2 spectroscopy}
\label{VLT/FORS2}

The spectrum of \psr was obtained in 2004 November--December and 2005 February  during several
observational runs\footnote{ESO program 074.D-0512A, PI R. Mennickent} using the VLT/UT1 telescope in a service mode (see
Table~\ref{t:logVLT}). The FORS2 instrument was used in a long slit spectroscopic setup with the GRIS\_300V grating 
and the $GG435$ filter, which cover the wavelength interval of about 4300-9600 \AA\ and provide a medium spectral
resolution of 3.35 \AA\ pixel$^{-1}$.  
The slit width was 1\farcs0 and its position angle was selected in such a way so it covered 
several nearby stars for the accurate  
astrometric referencing and wavelength/flux calibration (see. Fig.~\ref{fig:GTCslit}). Eighteen
1400~s science spectroscopic exposures were taken
with a total exposure time of 25.2~ks at a mean seeing of 0\farcs6.  
Standard reference frames (biases, darks, flatfields, lamps) were obtained  
in each observational run, while the slit and slitless observations of the spectrophotometric
standards (Feige110, LTT3218 and LTT1788) for the flux calibration were carried out in separate runs during the same nights.
A combination of the {\sc eso-midas} and {\sc iraf} packages was used for standard CCD data reduction, cosmic-ray
track removing, spectra extraction and the subsequent data analysis.

\psr~is a faint target with $R\approx$24.6, which is at the limit of the VLT spectroscopic
capability. Nevertheless, excellent seeing conditions allowed us to resolve its spectrum even on 
individual exposures, albeit with a low S/N ratio. The individual exposures were co-added.  
The spectrum  was then extracted with a 3 pixel wide extraction slit (0.2 arcsec pixel$^{-1}$) centred on the
pulsar. The backgrounds were extracted with  6 pixel wide slits centred above and below the centre of the
pulsar spectrum. The correction factor for the PSF and sensitivity function were obtained from the Feige110
standard  observations. The  S/N of the resulting  spectrum   was about 4 (per pixel) in the 4450-5500 \AA\ range 
and declined to $\sim 1$  near/above 8000 \AA,  due to higher
sky backgrounds  and a  drop in sensitivity  towards longer wavelengths. 
We binned the spectral flux in 20 pixel bins  (67\AA) to get S/N  near/above 15 and 4, 
respectively,  making the flux accuracy to be comparable with  that of available
photometric data. The resulting optical VLT spectrum of \psr is presented  in the top panel of 
Fig.~\ref{fig:PSRSpecRaw}.  It is in a good agreement with the  photometric data points. 
In general, it can be described by a single PL,  $F_{\nu} \sim \nu^{-\alpha} $, with a spectral 
index $\alpha=0.6(1)$.  We found no  evidence of any significant  narrow  (emission or absorption)
spectral features. An apparently large  spectral variation in the  flux density  at $\log \nu=
14.55 - 14.65$~Hz is due to  the low S/N in this range.    Broad weak  features could exist   in a
range of  $\log \nu= 14.63 - 14.82$~Hz, however, their significance cannot be  claimed
confidently.

\section{HST optical narrow-band photometry and  UV data; Subaru and BTA broad-band optical
photometry}
\label{HST/ACS}

\subsection{HST/ACS/WFC narrow-band photometry}

The {\it HST/ACS/WFC} observations of the pulsar field were obtained on 2005 December 3 and 12
using the $FR459M$, $FR647M$ and $FR914M$  filters with 11 central wavelength positions. We extracted the data from 
the {\it HST} archive and re-reduced them to  consistently compare with 
the narrow-band GTC photometry and the VLT spectroscopy. This analysis is important as based on these data 
\citet{2011ApJ...743...38D} reported on the presence of absorption/emission features in the spectrum of the pulsar, 
while we find no  such features in our VLT spectrum.    The log of the {\it HST} observations and results of our photometry
are given in Table~\ref{t:logHST}.  

We repeated the flux measurements of the pulsar in all available ACS/WFC bandpasses. 
For aperture photometry, the circular apertures with radii of 2, 3, and 4 pixels were used
(1~pixel = 0\farcs05). The aperture centre was chosen by fitting the pulsar profile with a Gaussian in each band (see
Fig~\ref{fig:ACS}). The background was measured using the annulus with respective inner and outer radii $r_{in} = 9$
and $r_{out}=14$ pixels for all apertures.  Following the {\it HST/ACS/WFC} user
manual\footnote{https://hst-docs.stsci.edu/display/ACSDHB/5.1+Photometry}, the  source fluxes
measured in count-rates (cps) for each aperture  
were corrected to an aperture of 0\farcs5  using photometry of bright unsaturated nearby star-like
objects in the ACS FoV. These corrected fluxes for different initial apertures showed a
considerable scattering  demonstrating   correction uncertainties, which were comparable to 
statistical count-rate uncertainties. They were then averaged and the final flux errors were derived accounting 
for the statistical and aperture correction uncertainties. 
After that, we  performed corrections for the PSF from the  0\farcs5 aperture to the nominal
`infinite' aperture of 5\farcs5 and for the charge transfer efficiency (CTE) using  
the correction factors from \citet{2005PASP..117.1049S} and \citet{2009acs..rept....1C}. 
The fluxes in count-rates were transformed into $\mu$Jy using  PHOTFLAN key-words from image
headers. The factors, key-words, and the log of our measurements are given in Table~\ref{t:logHST}.
Generally, our results are in agreement with those reported by \citet{2011ApJ...743...38D}. 
Nevertheless, we reduced the scattering of deviations from the power-law fit of the optical
spectrum that was visible in previous flux measurements. In addition, we found  that the fluxes in the $FR459M$ band at 
$\lambda$=4028.805\AA~($\log(\nu_{CW})=14.872$~ Hz) and 4400.793\AA~($\log (\nu_{CW})=14.834$~Hz) were   slightly 
overestimated   due to the contribution of an extended background structure, likely to be a cosmic-ray track, 
located at the pulsar position.  The revised photometric results are shown in Fig.~\ref{fig:PSRSpecRaw}. As seen, 
they are consistent with the VLT spectroscopy and do not show any significant spectral feature.

\subsection{HST UV, BTA and Subaru optical data}
\label{BTAdata}

 The HST UV data on the pulsar were obtained by \citet{2005A&A...440..693S} and \citet{2011ApJ...743...38D} using
the  STIS NUV-MAMA and COS-FUV instruments, respectively. These data were found to be self-consistent  \citep[see fig.4 therein]{2011ApJ...743...38D}. We also included in our analysis the published results of the broad-band optical photometry of the pulsar obtained with the Subaru \citep{2006A&A...448..313S}, and BTA \citep{2001A&A...370.1004K} telescopes.
 These data were obtained by some authors of this 
 study and have been revised  and checked independently several 
 times in previous publications. 
Magnitudes of secondary photometric standards in the pulsar field from \citet{1998A&A...333..547K}, used for calibration of different sets of optical data,   agree with those presented 
in the Pan-STARRS \citep{2016arXiv161205560C} and SDSS \citep{2017AJ....154...28B} catalogues.
Therefore, we included in our analysis all these data on the pulsar in their original form.  
They are also presented in Fig.~\ref{fig:PSRSpecRaw}.

%


\section{X-ray data}
\label{sec:Xrays}

To perform multi-wavelength analysis of the pulsar spectrum from the near-IR-optical through
X-rays, we selected the best-quality published X-ray data sets obtained with \textit{ XMM-Newton},
and \textit{NuSTAR} observatories  in a photon energy range of 0.15--20~keV (for references see
Table~\ref{tab:X-ray}). We re-reduced the archival X-ray data and extracted time integrated pulsar spectra for fitting
them by various models simultaneously  with the long wavelength data. Below we briefly describe
each of the sets. 

The \textit{XMM-Newton} observations were carried out on 2015 September 19 (observation  ID 0762890101)
and 2019 October 14 (observation  { ID 0853000201}) with total exposure times $\approx 130$ ks
and $\approx 72$ ks, respectively.  The latter data set has not been published yet, but it adds a considerable number of source counts to those of the first set, thus allowing one to improve the spectral analysis results. 
 In the observations, the EPIC-MOS cameras were operated in the
Timing Mode with the THIN filter setting, and the EPIC-{\it pn} camera was operated in the Small
Window Mode with the THIN filter. The {\sc sas} v.17.0.0 software was used to process the data. 
In our analysis, we used only the data from the {\it pn} camera. 
We selected single and double pixel events (PATTERN $<=$ 4).  
Considering light curves we removed periods of background flares.   
After that, the science  useful {\it pn} exposures became  84 ks and 71 ks for the first and second sets, respectively. 
We applied the circular apertures with radii of 37\farcs5 and 15\farcs0 for energy ranges 
of 0.15 -- 1.4 and 1.4 -- 7 keV 
and extracted the pulsar spectra using {\sc evselect} tool. The apertures were chosen to
increase signal-to-noise of the pulsar spectra (see, e.g. \citet{2018XMM}). Background was taken
from a region free from any sources. We then used {\sc sas} tasks {\sc rmfgen} and {\sc arfgen} to
generate the redistribution matrix and ancillary response files for the source and background.
We analysed the  spectra in the  0.3--7~keV range, which is the nominal range for the {\it pn}. 
In addition, we checked {\it pn} spectra in the 0.15--7~keV range, despite the fact that the 0.15--0.3 keV range
is not well calibrated\footnote{https://xmm-tools.cosmos.esa.int/external/ 
xmm\_user\_support/documentation/uhb/epicintbkgd.html or
http://xmm2.esac.esa.int/docs/documents/CAL-TN-0018.pdf}.

The \textit{NuSTAR} telescope observed the pulsar
in 2015 with FPMA and FPMB detectors (observation  ID 40101004002).
We retrieved    the  \textit{NuSTAR}  data   from the archive and reduced them in a similar way
as \citet{2018XMM}. \textit{NuSTAR} spectra of the pulsar were analysed in the 3--20 keV range. 

\begin{table*}
\caption{Results of the spectral fits of the time integrated emission from PSR B0656$+$14.}  
\label{tab:Fit:M1} 
\begin{tabular}{lccc|||ccc} 
\hline
Model                             & G2BB + PL                  & G2BB + PL                 & G2BB + BKPL              & G2BB + BKPL                    & 2G2BB + BKPL               \\
\hline
spectral                     & X-rays                      & nIR-opt-UV             & nIR-opt-UV             & nIR-opt-UV                           & nIR-opt-UV             \\ 
range& in 0.3$-$20~keV  & and X-rays in & and X-rays in  & and X-rays in  & and X-rays in  \\
& & 0.3$-$20~keV & 0.3$-$20~keV & 0.15$-$20~keV  & 0.15$-$20~keV \\
\hline 
fit ID                            & {\it  N1}                  &{\it  N1o}                 & {\it  N2}                & {\it  N3}                      & {\it  N4}                  \\
\hline 
\hline
Parameter                         &                            &                           &                          &                                &                            \\     \hline
$N_{\rm H}$ ($10^{20}$ cm$^{-2}$) & $4.2^{+1.0}_{-0.9}$        & $3.4^{+0.3}_{-0.2}$       & $3.1^{+0.4}_{-0.2}$      & $2.00^{+0.10}_{-0.10}$         &   $1.96^{+0.20}_{-0.28}$   \\
$E(B-V)$                           & $0.047^{+0.011}_{-0.010}$  & $0.038^{+0.003}_{-0.002}$ & $0.035^{+0.004}_{-0.002}$& $0.022^{+0.001}_{-0.005}$     & $0.022^{+0.002}_{-0.003}$  \\     \hline 
$E_{\rm c}$ (keV)                 & $0.53^{+0.02}_{-0.03}$     & $0.547^{+0.007}_{-0.006}$ & $0.552^{+0.006}_{-0.006}$& $0.570^{+0.006}_{-0.007}$      & $0.548^{+0.007}_{-0.008}$   \\
$\sigma$ (keV)	                  &  $0.13^{+0.02}_{-0.03}$    & $0.117^{+0.029}_{-0.018}$ & $0.110^{+0.017}_{-0.017}$& $0.068^{+0.008}_{-0.010}$      & $0.122^{+0.025}_{-0.021}$   \\
$s$                               &  $0.11^{+0.09}_{-0.06}$    & $0.09^{+0.05}_{-0.03}$    & $0.076^{+0.038}_{-0.026}$& $0.026^{+0.006}_{-0.05}$       & $0.093^{+0.058}_{-0.043}$   \\
\hline 
$E2_{\rm c}$ (keV)	              &          -                 &            -              &             -            &        -                       & $0.300^{+0.007}_{-0.018}$   \\
$\sigma2$ (keV)                   &          -                 &            -              &             -            &        -                       & $0.035^{+0.019}_{-0.016}$   \\
$s2$                              &          -                 &            -              &             -            &        -                       & $0.019^{+0.018}_{-0.005}$   \\ \hline 
BB1$_{norm}$ ($10^{5})$           & $2.42^{+1.91}_{-0.82}$     & $1.67^{+0.07}_{-0.09}$    & $1.52^{+0.11}_{-0.12}$   & $1.17^{+0.09}_{-0.10}$         & $1.26^{+0.10}_{-0.08}$      \\
$R_{\rm BB1}$ (km)                & $14^{+5}_{-2.5}$           & $11.76^{+0.25}_{-0.43}$   & $11.22^{+0.41}_{-0.46}$  & $9.87^{+0.37}_{-0.47}$         & $10.20^{+0.40}_{-0.32}$     \\
$kT_{\rm BB1}$ (eV)               & $65^{+3}_{-3}$             & $68^{+1}_{-1}$            & $68^{+1}_{-1}$           & $67^{+1}_{-1}$                 & $69^{+2}_{-2}$              \\    \hline
BB2$_{norm}$                      & $756^{+423}_{-260}$        & $541^{+194}_{-155}$       & $662^{+160}_{-204}$      &  $920^{+166}_{-172}$           &  $600^{+255}_{-223}$        \\ 
$R_{\rm BB2}$ (km)                & $0.80^{+0.20}_{-0.16}$     & $0.67^{+0.11}_{-0.11}$    & $0.74^{+0.09}_{-0.12}$   & $0.87^{+0.08}_{-0.09}$         & $0.71^{+0.13}_{-0.14}$      \\
$kT_{\rm BB2}$ (eV)               & $129^{+6}_{-5}$            & $134^{+6}_{-4}$           &  $131^{+6}_{-3}$         & $127^{+3}_{-3}$                & $132^{+6}_{-5}$             \\    \hline
$\Gamma$                          & $1.79^{+0.13}_{-0.16}$     & $1.54^{+0.01}_{-0.01}$    & $1.74^{+0.12}_{-0.11}$   & $1.82^{+0.13}_{-0.11}$         & $1.72^{+0.14}_{-0.09}$      \\ 
$\Gamma_{opt}$                    & --                         & $=\Gamma$                 &  $1.44^{+0.05}_{-0.06}$  &  $1.35^{+0.06}_{-0.05}$        &  $1.45^{+0.05}_{-0.06}$     \\
PL$_{norm}$ $(10^{-5})$           & $2.53^{+0.46}_{-0.41}$     & $1.97^{+0.11}_{-0.09}$    & $3.82^{+1.83}_{-1.15}$   &  $6.84^{+2.38}_{-2.04}$        &  $3.47^{+1.60}_{-1.05}$     \\
E$_{break}$ $(keV)$               &          --                &           --              & $0.130^{+0.470}_{-0.125}$& $0.120^{+0.201}_{-0.088}$      & $0.127^{+0.656}_{-0.101}$   \\
\hline 
$ln(Z)$                           &   -307                     & -390                      & -380 & -445  & -394 \\
n                                 &   548                      &  620                      & 620  & 672   & 672 \\
k                                 &   12                       &  12                       &  14 &  14    &  17  \\
BIC                               &  690                       & 857                       & 850 & 981    & 899 \\
\hline 
\end{tabular}  
\begin{tablenotes}  
\item $E_c$ is   the Gaussian absorption line central energy, $\sigma$ and $s$ are  its width and
strength, respectively.
BB$_{\mathrm{norm}}=R_{BB}^2/D^2_{10}$ is the blackbody normalisation constant,  where $D_{10}$ is
the distance to the  source is in units of 10 kpc (it is fixed at 288 pc),  
$R_\mathrm{{BB}}$ is the radius in km of the BB emitting area, $kT_\mathrm{{BB}}$ is the BB
temperature. $\Gamma$ is photon index of the PL component and    
PL$_\mathrm{norm}$ is a its normalisation  in units of photons keV$^{-1}$ cm$^{-2}$ s$^{-1}$ at 
1~keV, E$_{break}$ is the energy of the PL spectral break.   
The errors are  given at 90\% credible interval. $n$ and  $k$ are numbers of data points and
varied parameters, respectively. $lnZ$ is the maximum log-likelihood value and BIC =
k$\times$ln(n)-2$\times$ln(L) + is Bayesian information criterion. 
\end{tablenotes} 

\end{table*}


\section{Multiwavelength spectral analysis}
\label{sec:msf}

\subsection{Spectral fitting setup}

We used the {\sc xspec} v.12.9.0 \citep{Arnaud:1996aa} tool for fitting  all the data from the
near-IR to hard X-rays by theoretical models describing thermal and nonthermal emission of the
pulsar. The near-IR--UV data were incorporated into XSPEC using 
the {\sc ftflx2xsp} tool \citep{1995ASPC...77..367B}. \textit{XMM-Newton} X-ray spectra 
were grouped only in the energy range of 1.4 -- 7 keV  to ensure signal-to-noise of 4 per 
energy bin and used in our fit simultaneously.  The floating cross-normalisation was thawed  
for each data set while it was fixed at 1 for the \textit{ XMM- Newton} data. 

For the X-ray photoelectric absorption, we used the {\sc xspec}   model {\sc tbabs} with  abundances 
{\sc wilm}  \citep{WILM} and cross-sections {\sc bcmc} \citep{Balucinska-Church:1992aa}.  
To account for the interstellar extinction in the optical--UV, we used the XSPEC  model 
{\sc redden} utilizing the extinction law from \citet{Cardelli:1989aa}. Its only parameter, 
the optical  colour excess $E(B-V)$, was linked to  the X-ray absorbing column density 
$N_{\mathrm H}$ using the relation from \citet{nh-ebv2016}, 
$E(B-V)=N_{\mathrm H}/8.9 \times$10$^{21}$~cm$^{-2}$. 
The optical transmission  {\sc redden}  is set to unity shortward 
of the Lyman limit, i.e. it does not affect the X-ray data. This is incorrect physically 
but does allow the model to be used in a combination with the X-ray photoelectric absorption
model, which in turn  does not directly affect the optical-UV data.

We implemented the fitting by a Markov chain Monte-Carlo
(MCMC) sampling procedure assuming a uniform prior distribution for model parameters. 
We employed the affine-invariant MCMC sampler developed by \citet{Goodman:2010aa} and 
implemented in a {\sc python} package {\sc emcee} by \citet{Foreman-Mackey:2013aa}.
About 100 walkers and 15000 steps were typically enough to ensure  fit convergences. 
Using the sampled posterior distribution, we obtained best-fit estimates and 
credible intervals of the model parameters. 

\subsection{Spectral fitting results} 

Following  recent X-ray studies (see Sect.~\ref{intro}), we first applied the absorbed 
G2BB$+$PL spectral  model. It includes 2 BB components   describing thermal emission
from the bulk of the surface of the cooling NS (BB1) and its hot polar caps (BB2), the PL
component from the magnetosphere of the pulsar and  the  Gaussian absorption line 
(G;{\sc xspec} model {\sc gabs})  to account for the X-ray spectral feature reported by
\citet{2018XMM}. Using the X-ray data alone  we checked that the model  provides an acceptable 
fit to the data in a range of 0.3--20 keV  with parameters (Table~\ref{tab:Fit:M1}, the fit ID 
{\it N1}) generally consistent with those obtained by these authors 
using only the first \textit{XMM-Newton} data set in the 0.3-7 keV range. 
Adding the second \textit{XMM-Newton}  set and the \textit{NuStar} data results only in a
marginal increase of the spectral index of the PL component ($\Gamma=1.79^{+0.13}_{-0.16}$ {\sl
vs} $1.7\pm0.1$) and the absorbing column density ($N_H=4.2^{+1.0}_{-0.9}$~cm$^{-2}$  {\sl vs} 
$3.0^{+0.7}_{-0.9}$~cm$^{-2}$) and a decrease of the radius 
of the pulsar `hot spot' ($R_{BB2}\approx800$~m {\sl vs} $\approx1000$~m).

\begin{figure*}
\setlength{\unitlength}{1mm}
\resizebox{15.cm}{!}{
\begin{picture}(130,130)(0,0)
\put (0,65){{\includegraphics[width=13cm,bb=30 260 560 522,  clip=]{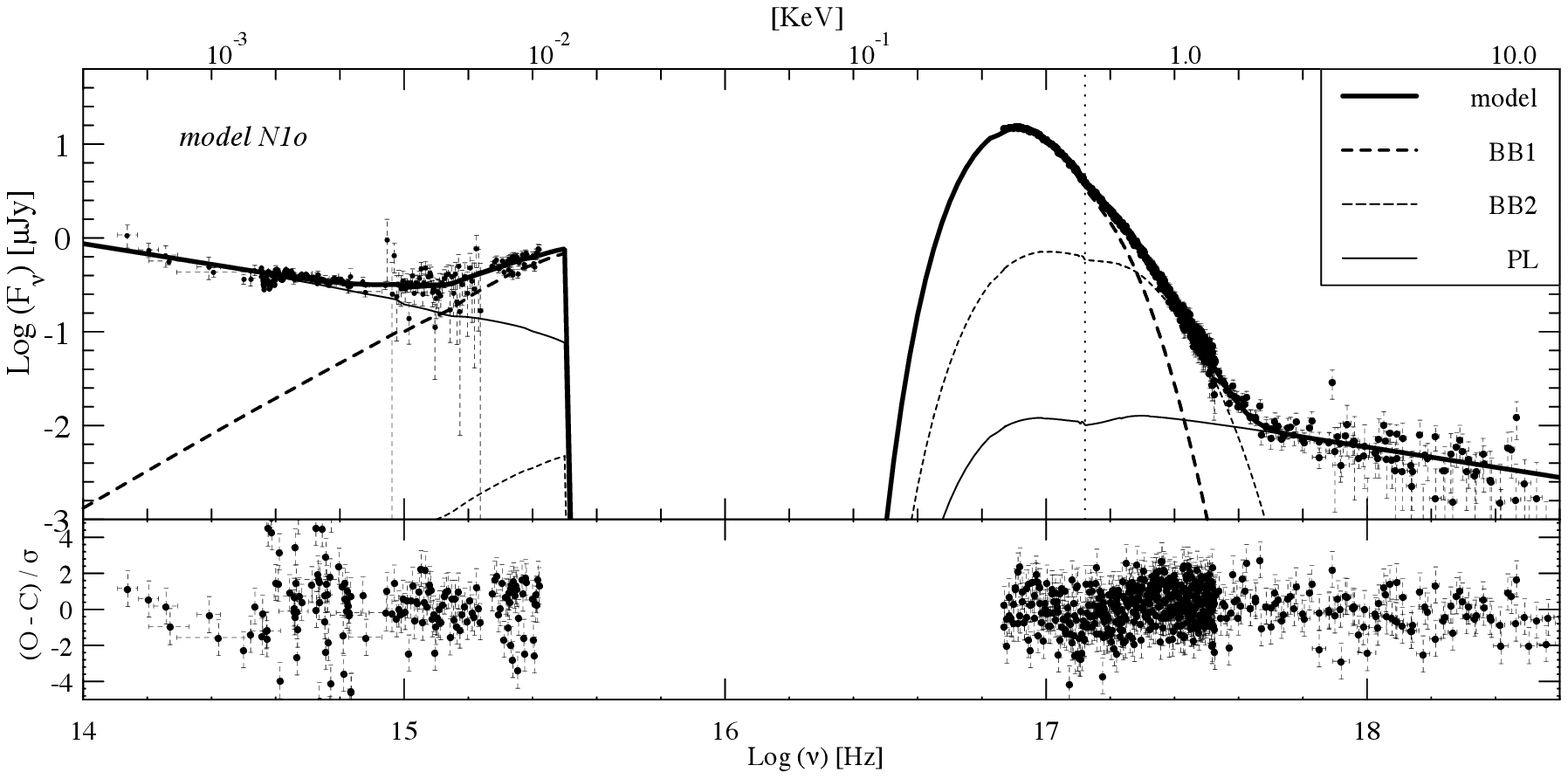}}}
\put (0,0){{\includegraphics[width=13cm,bb=30 260 560 522,  clip=]{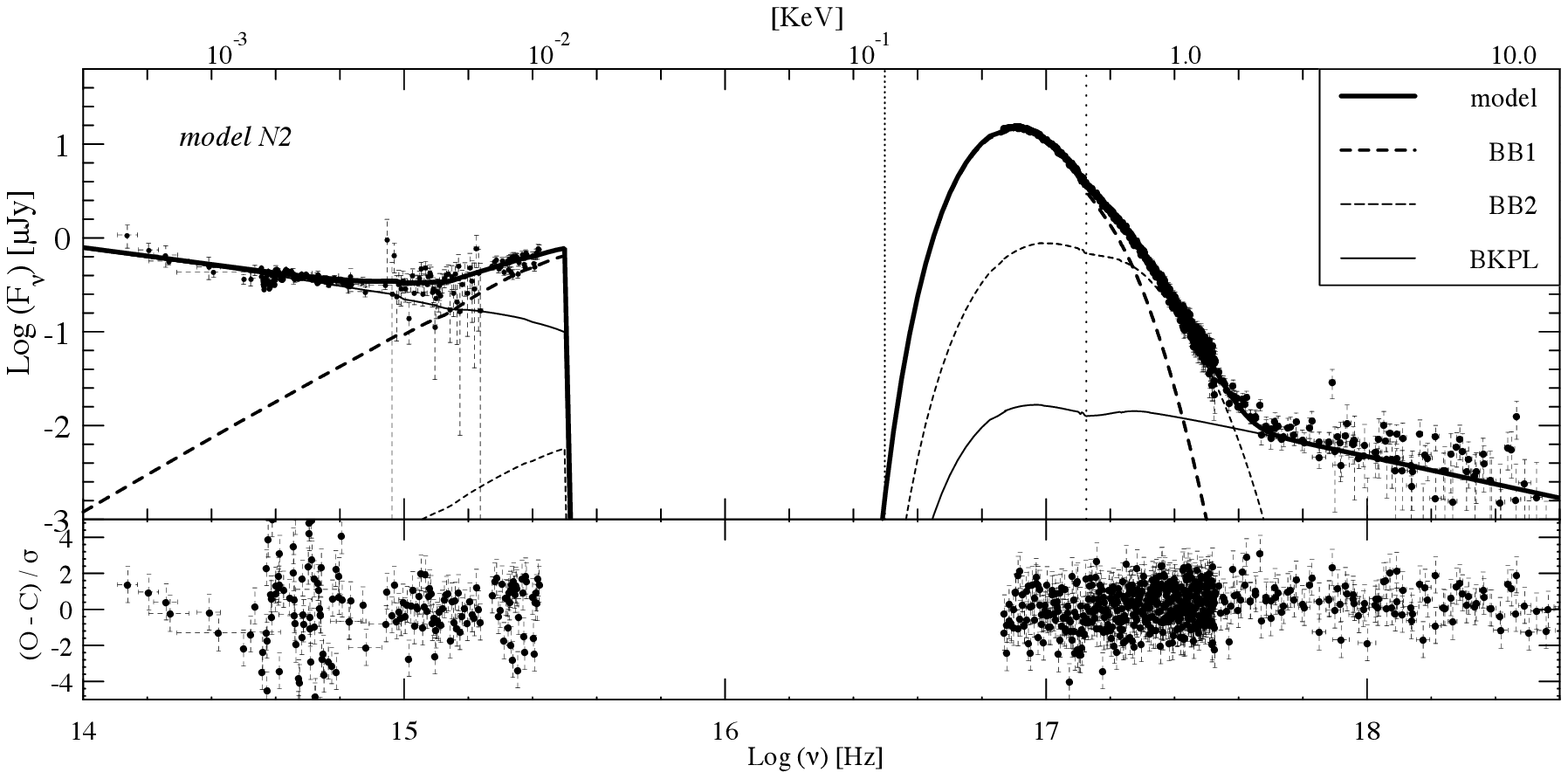}}}
\end{picture}}
\caption{Observed multiwavelength time-integrated spectrum of the pulsar from the IR through
X-rays fitted with the absorbed G2BB+Pl ({\sl top }) 
and G2BB+BKPL ({\sl bottom}) models. The VLT optical spectrum is shown without errors to simplify
the plot. The best fit models with their components, BB1, BB2, PL/BKPL, are shown by different
type lines as notified in legends. The best fit positions of the absorption line and the
nonthermal component spectral break (BKPL case) are  marked by  vertical dotted lines. The fit
residuals expressed in terms of the difference between the observed data (O) and the calculated
model (C) divided by the data error in each spectral point are shown in the {\sl low sub-panels}
of  both plots.}
\label{fig:AllFit1}
\end{figure*}

\begin{figure*}
\setlength{\unitlength}{1mm}
\resizebox{15.cm}{!}{
\begin{picture}(130,130)(0,0)
\put (0,65){{\includegraphics[width=13cm,bb=30 260 560 522,  clip=]{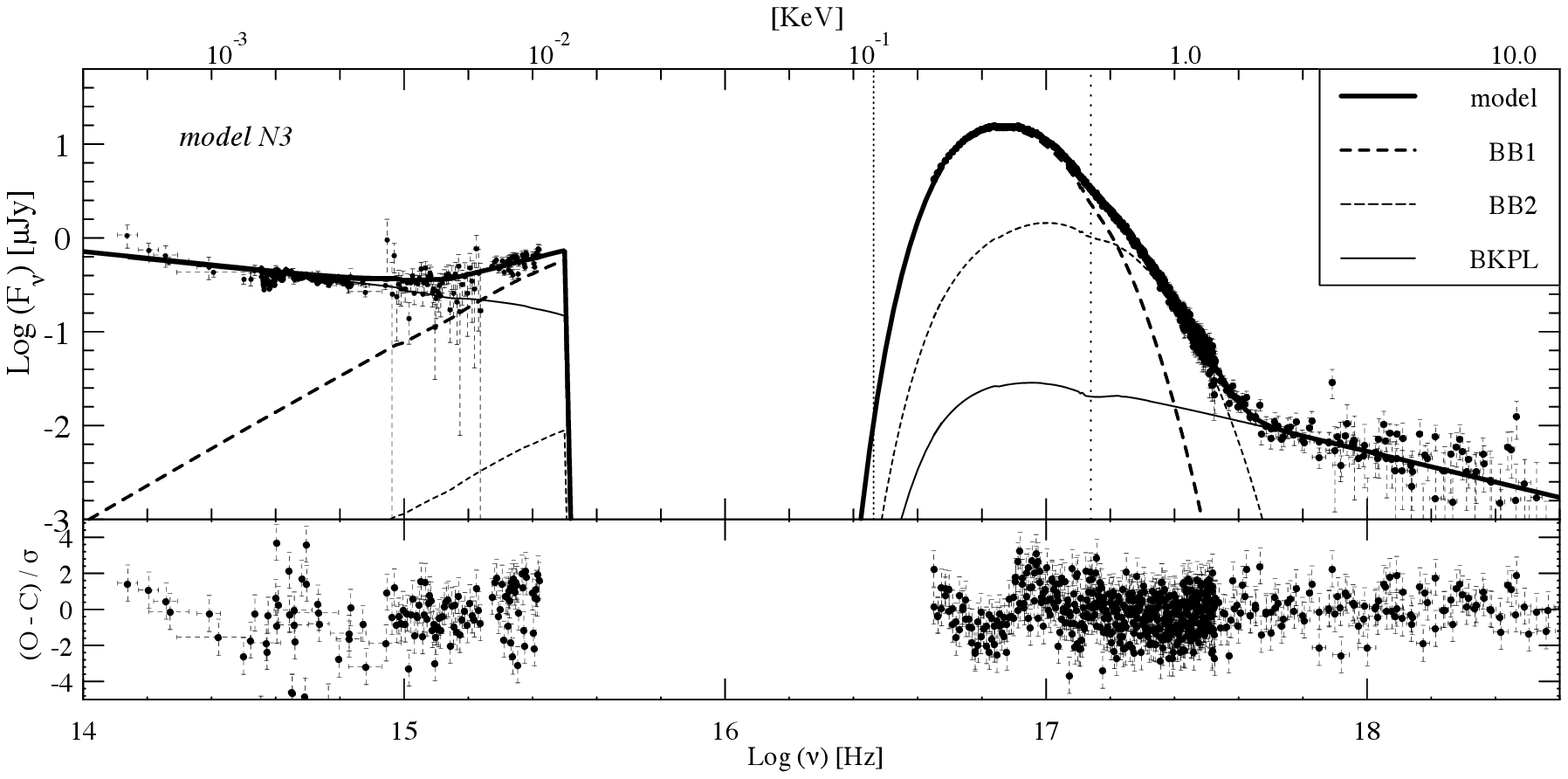}}}
\put (0,0){{\includegraphics[width=13cm,bb=30 260 560 522,  clip=]{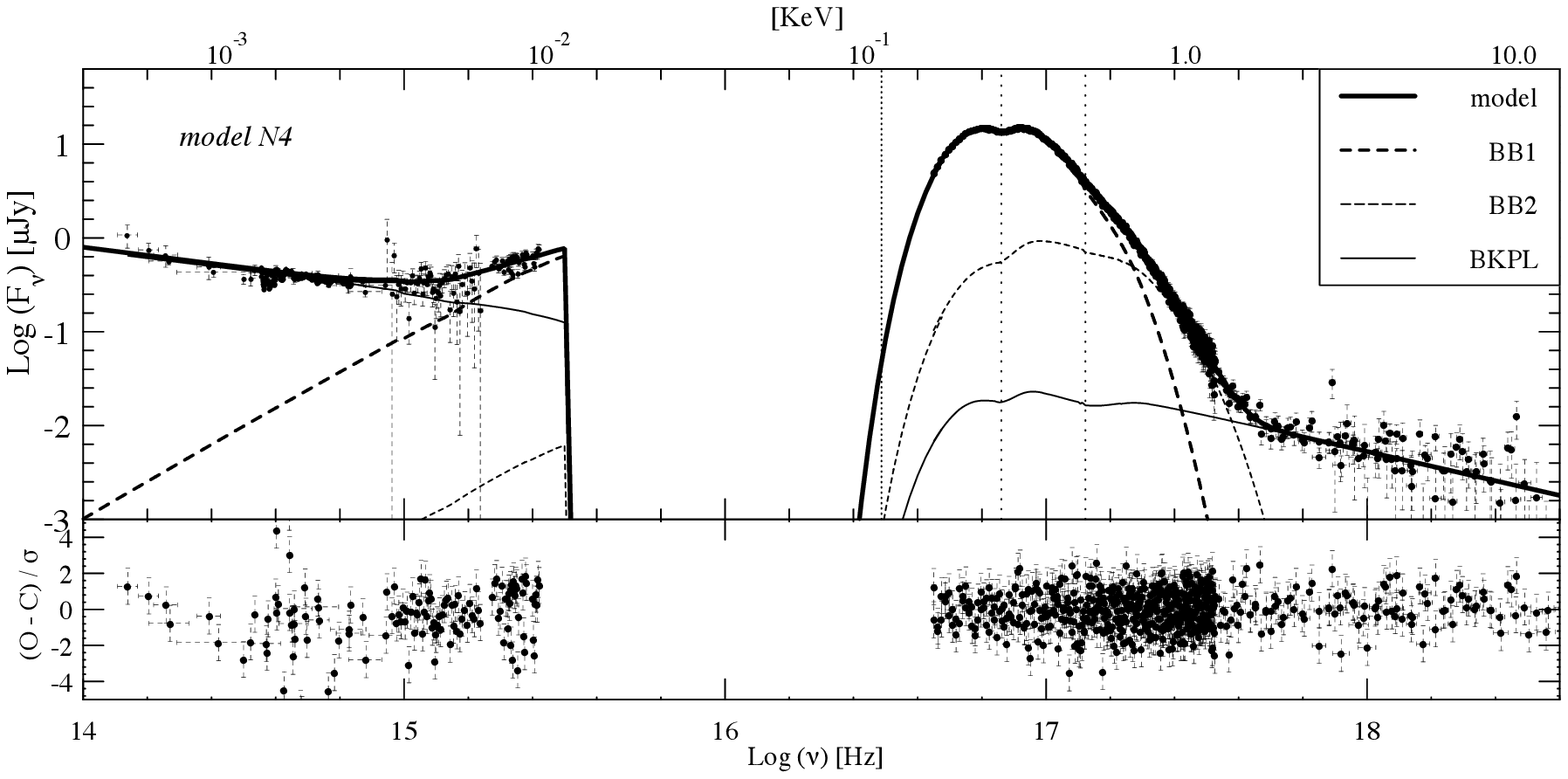}}}
\end{picture}}
\caption{The same as Figure~\ref{fig:AllFit1} but including the data from the 0.16--0.3 keV range and 
using the G2BB+BPKL ({\sl top}) and 2G2BB+BPKL ({\sl bottom}) models to fit the spectrum. 
As seen, the addition of the second Gaussian absorption line into the model excludes prominent
fit residuals seen in the soft part of the X-ray spectrum in the {\sl top} panel where the model
contains only one Gaussian line.  
}
\label{fig:AllFit2}
\end{figure*}

Then, we compared the long wavelength extrapolation  of the  fit {\it N1} with the observed optical-UV data. 
This is presented in the lower panel of Fig.~\ref{fig:PSRSpecRaw} where  absorbed  BB1 and PL  contributions and their 
sum  are shown by low, middle and upper dashed lines. The hot BB2 component has a negligible contribution 
in this range and is not shown. As seen, the extrapolation overshoots the near-IR-optical data by a factor of five. 
Overshooting the UV data, where the thermal component starts to dominate, is less significant
but noticeable (about 1$\sigma$). Thus the model with parameters which fit well  the X-ray
spectrum alone cannot describe the data in the near-IR-optical range. 

Therefore, as a next step, we fitted the X-ray and near-IR-optical-UV data simultaneously  
using the same model. The output of the fit is presented in the Table~\ref{tab:Fit:M1}
(the fit ID {\it N1o}). For the near-IR-optical-UV domain, 
its results are demonstrated in the lower panel of Fig.~\ref{fig:PSRSpecRaw} 
where the absorbed BB1 and PL  spectral components and their sum are shown by solid lines. 
The unfolded absorbed best fit  model with all its components
together with the whole set of the multiwavelength data are presented 
the in top panel of Fig.~\ref{fig:AllFit1}. According to the fit residuals 
(the low sub-panel of this plot), the model  describes well  
the data  in the whole observing range.  Parameters of the thermal components  
are consistent with those  obtained using the X-ray data alone ({\it N1} fit).   
However,  their uncertainties, as well as the uncertainty of $N_H$, 
are much smaller.  A significant difference between the two cases demonstrates only  the PL
component, whose spectrum becomes  less steeper when we include the long wavelength data.
The  photon index of the nonthermal  component $\Gamma=1.54\pm 0.05$ obtained from the 
 fit {\it N1o} is consistent with  the spectral index $\alpha=\Gamma-1=0.51\pm 0.03 $ 
derived from the PL fit of the  optical spectrum alone dereddened with the  
$E(B-V)\approx 0.038$ provided by the {\it N1o} fit. On the other hand, this  component  
also  describes apparently well the  nonthermal spectral tail seen in hard X-rays. 
 Although the  fit  {\it N1o } is formally acceptable, implying a single PL for the 
 optical and X-ray data,  the large difference between the  
  extrapolation of the  fit {\it N1  } and the optical-IR data mentioned above 
 indicates that the PL component possibly has  a spectral break.

To investigate whether such a break exists in the PSR B0656$+$14 spectrum and to 
estimate its parameters, we  tried the G2BB$+$BKPL model, where  the BKPL is the broken PL component. 
The rest of the components are the same as in the previous model. The  fit, below referred as {\it N2},
describes the multiwavelength data  almost equally well  as the {\it N1o} fit (Table~\ref{tab:Fit:M1} and the bottom
panel of Fig.~\ref{fig:AllFit1}). However, using Bayesian information criteria \citep[BIC;][]{claeskens_hjort_2008}, we 
inferred that {\it N2} has a smaller BIC than {\it N1o}, making the former more preferable 
for the same data sets (see Table~\ref{tab:Fit:M1}). Specifically, the obtained $\Delta BIC=7$
implies {\it substantial  evidence  against} {\it N1o}. In addition,  we evaluated the Bayesian evidence $Z$
(also shown in Table~\ref{tab:Fit:M1}) and calculated the Bayesfactor \citep{Goodman1999a,Goodman1999b} 
for {\it N2} over {\it N1o},  $\Delta lnZ=ln(Z_{N2}/Z_{N1o})=10$.
Assuming equal prior odds for the two models, the posterior odds  ratio  for  the models  
with the broken power law over the single power law is $Z_{N2}/Z_{N1o}\approx 22000$. 
This implies that the G2BB+BKPL model is much more probable than the G2BB+PL one,  
which strongly favours   the presence of the spectral break in the nonthermal 
spectral component of the pulsar. Namely, its spectrum  becomes steeper above the break, 
located at the photon energy between 0.005 -- 0.6~keV, with  the slope  described by  $\Gamma=1.74 \pm 0.12$. 
The parameters of both thermal components, $E(B-V)$ and $N_H$ remain practically the same 
as for the  fit \textit{N1o}.  Ultimately, the G2BB$+$BKPL model combining  thermal emission from the bulk 
of the surface of the cooling NS,  its hot polar caps,  and the BKPL emission component of 
the magnetosphere origin  appears to represent the most robust description of the current
time-integrated multiwavelength spectrum of the pulsar.

For completeness, we also included   the 0.15 -- 0.3 keV part of the 
{\textit XMM-Newton}/EPIC-pn spectrum of  the pulsar so far ignored in the analyses due to
possible calibration issues partially related to substantial  internal background increase of
the pn instrument below 0.3  keV (see Sect.~6). Applying the G2BB+BKPL model to fit the extended
data set (the fit ID \textit{N3}) shows characteristic fit residuals in this part of the spectrum
indicating a possible presence of a second absorption line  (see {\sl top panel} of
Figure~\ref{fig:AllFit2}). Indeed, fitting with the 2G2BB+BKPL model,  
which includes two Gaussian absorption lines, excludes the residuals ({\sl bottom panel} of
Figure~\ref{fig:AllFit2}) and allows us to derive the line  parameters 
(the fit ID  \textit{N4} in Table~\ref{tab:Fit:M1}).  It is interesting, that the energy of the
second line centre derived from the fit,   $E2_c\approx 0.3$~keV,  is remarkably 
close to a half
of the central energy $E_c\approx0.6$~keV of the first line, implying that they could represent
cyclotron absorption lines created in the magnetised plasma of the pulsar.   
The  Bayesian evidence and information criteria $Z$ and BIC  of the  fit {\it N3} are significantly
larger and smaller, respectively,  than those in the case the single Gaussian line model (the fit
{\it N3}), formally strongly favoring the presence of the second line. 
We  also note that the interstellar absorption and radius of the cold thermal component become 
 slightly smaller when we include the extended X-ray data set. However, all these issues,
 including  the presence of the second line should  be considered with caution. A reliable 
 solution of the EPIC-pn  low energy calibration problems and independent detection of the second 
 line with \textit{eROSITA } are required to confirm them.

\section{Discussion}
\label{sec:disc}

Our first spectroscopic, narrow-band  optical, and broad-band near-IR observations 
of \psr with the VLT and the GTC together with the re-analysed  archival data obtained with 
the {\it HST} and the optical-UV data published since its first identification in the optical 
about 26 years ago by \citet{1994ApJ...422L..87C}   represent  
the most reliable and  up-to-date complete information on  the spectral energy distribution (SED) 
of the pulsar emission in the near-IR-optical-UV range.  The near-IR-optical part of the  SED can be described by 
a single PL with the  spectral index $\alpha_{\nu} \approx 0.6$ (see Fig.\ref{fig:PSRSpecRaw}, top) thus confirming 
its pulsar magnetosphere origin. The obtained data exclude the presence of any strong narrow spectral lines.  
There are some hints of weak broad spectral features between 4500\AA\ and 7500\AA\ but their
presence can be confirmed only by observations with a higher S/N. Our first observation of the
pulsar in the K$_s$ band might allude to  the presence of the IR excess 
over the  PL optical SED (see Fig.~\ref{fig:PSRSpecRaw}). However, its current significance 
is only about 1$\sigma$. The available {\it Spitzer} data do 
not allow us to confirm it due to the strong contamination of the 
pulsar flux by a distant background galaxy located north-west of the 
pulsar in its nearest vicinity (see Fig.~\ref{fig:GTCir}, bottom). High spatial resolution 
near- and mid-IR observations  with large ground-based telescopes equipped  by AO systems or
with the {\it JWST} space telescope are necessary to resolve the pulsar from the galaxy. That would 
be intriguing, as  a firm flux increase towards the IR was confirmed so far only for the young 
PSR B0540$-$69 \citep{2012A&A...544A.100M}.

After including two archival data sets obtained 
with \textit{XMM-Newton} and observations with \textit{NuSTAR} into our X-ray spectral analysis, we confirmed the conclusion of
\citet{2018XMM} based on the single \textit{XMM-Newton} data set  on the presence of the
absorption line near $\approx 0.6$~keV in the time-integrated 
X-ray spectrum of the pulsar whose  X-ray  emission in continuum is best described by two
blackbody and  power-law spectral components. On the other hand, 
analysing simultaneously the near-IR-optical-UV SED  and X-ray data, we conclude  that 
the nonthermal emission component  cannot be described by a single power-law in the whole observed
range.    A flatter near-IR-optical spectrum with photon index  $\Gamma_{opt}$=1.44$\pm0.05$ 
steepens significantly to  $\Gamma_{X-ray}$=1.74$^{+0.12}_{-0.11}$  with  increasing  photon
energy  towards  X-rays suggesting  the spectral break between the two ranges. According to our
multiwavelength spectral fits the break is presumably located near $\approx 0.1$~keV. 
The presence of spectral breaks in the PL component between the optical and X-rays is 
typical for most pulsars detected in both spectral ranges \citep[e.g.,][and references
therein]{2006A&A...448..313S, Zharikov:2008aa, Kirichenko:2014aa}  while no multi-wavelength fits have been performed 
for other pulsars yet. 

 Including the near-IR-UV SED into the spectral fits together with the X-ray data allows us
 also to significantly better and self-consistently constrain the parameters of the thermal and 
 nonthermal emission components of the pulsar and   the interstellar absorption towards it in the
 optical and X-rays than it is possible using  the  data only in separate spectral ranges. For
 instance, the temperature as measured by a distant observer of the thermal component  from the
 bulk of the surface of the NS, $kT^{\infty}\approx68$~eV  
 ($T^{\infty}\approx7.9\times10^5$~K) is derived from the multiwavelength 
 fit  with the accuracy of 1 eV at the 90\% credibility level. The latter is a factor of four  better than it was
 obtained by \citet{2018XMM} using the  X-ray data alone.  Accounting for the gravitational redshift\footnote{
 $T=T^{\infty}/\sqrt{1-2.953M_{NS}(M_{\sun})/R_{NS}(km)}$}, this yields the NS surface temperature of
 $\approx 9.6\times 10^5$~K at $M_{NS}=1.4 M_{\sun}$ and $R_{NS}=13$~km.
 High quality  measurements of the effective surface temperatures of isolated cooling NSs are important for comparison 
 with theoretical 
 cooling curves of NSs and obtaining information on still poorly known properties of super-dense matter in their
 interiors \citep[e.g.,][]{2004yp}. Among about fifty NSs whose surface temperatures were estimated mainly based 
 on the X-ray data \citep{2020MNRASpalex} only for  a few stars the temperatures 
 are measured with a high accuracy comparable to that obtained here for PSR B0656$+$14. 
 Significantly  larger temperature errors, at least by a factor of ten,  for the rest NSs complicates
 the data comparison with the theory. As shown here, multiwavelength data can help to considerably
 decrease the uncertainties and facilitate the progress in this direction.    
 
 Based on the most robust G2BB+BKPL model  the bolometric thermal luminosity of the bulk of the 
 NS surface is  $\log(L_{BB1}^{\infty})=32.54\pm0.15$~erg s$^{-1}$ accounting for the temperature,
 radius and distance uncertainties.  The effective BB radius of the emitting area of the cool thermal component 
 BB1 in the G2BB+BKPL model, $11.22^{+0.41}_{-0.46}$ km, is translated to the  circumferential  
 radius\footnote{$R=R^{\infty}\sqrt{1-2.953M_{NS}(M_{\sun})/R_{NS}(km)}$} of $9.3\pm0.4$~km, accounting for the 
 gravitational light bending near the surface of the NS with $M_{NS}=1.4 M_{\sun}$ and $R_{NS}=13$~km. 
 In case if we accept that the emitting region of the BB1 component corresponds to the whole NS surface, 
 the radius of the pulsar becomes unrealistically small, only $R_{NS}=5.3$km for $M_{NS}=1.4 M_{\sun}$, 
 and slightly larger for the smaller pulsar mass.


The fact that the circumferential radius of \psr is  
smaller than the most  plausible NS radius  $R_{NS}=10.7-13.1$~km for a neutron star mass of 1.4M$_{\sun}$   
\citep[e.g.,][]{LattPrak2016, 2020ApJ...899..164H} shows that the temperature distribution over the NS surface is
likely to be non-uniform 
with a  gradient from the magnetic pole to equator, and its surface emission cannot be described by a simple BB model
with a single temperature. Such  models were considered by \citet{1995NYASA.759..291S} and \citet{possenti}  
to explain $\sim$20\%  pulsed emission fraction in 
soft X-rays. It is also in accord with the analysis of pulsations of the soft blackbody component and  phase 
resolved X-ray spectroscopy
performed by  \citet{2018XMM} showing that  $R_{BB1}$ can vary with the pulsar rotation 
phase from $\approx$6~km to $\approx$22~km in anti-phase  to the temperature.  
Therefore, the radius derived by us from the phase integrated spectra is only an effective radius of radiation from the NS  and does 
not directly correspond to the real physical size of the object.

BB models provide acceptable fits to the thermal components of the spectra of \psr, in contrast to  NS atmosphere 
models. This indicates that the observed UV and soft X-ray radiation come from a bare condensed surface of the NS.     
Indeed, \citet{2005ApJ...628..902V}  
showed that the bare condensed surface of a NS  produces a spectrum close to 
a diluted blackbody. 
Therefore,  
application of the bare surface model  assuming a plausible distribution of 
the temperature over the surface of the NS and accounting for the pulsar viewing geometry would be 
the next physically reasonable  step in the interpretation of the current and future multiwavelength data of \psr. 
This can provide the most robust constraints on the thermal emission parameters 
of the NS  based on the analyses of the phase resolved spectra and light curves   
of the pulsar in different spectral bands. $R_{NS}$ and $M_{NS}$  can be also  constrained this way as  
has been recently demonstrated for the isolated millisecond PSR J0030$+$0451 using \textit{NICER} data
\citep{2019ApJ...887L..24M,2019ApJ...887L..21R}.

On the other hand, the effective BB radius of the emitting area of the hot thermal component BB2 of about 740 m,  is converted  
to the intrinsic radius of 630 m under the same assumptions. This is about twice as large as the 
`canonical' polar cap radius of \psr, $R_{cp} = R_{NS} \sqrt{2 \pi R_{NS}/cP}$ $\approx$ 360~m. 
Therefore, using the BB model with a single temperature for this component should be also
considered with caution as the simulations predict non-uniform temperature distribution around the polar cap \citep{2002ApJ...568..862H} with a temperature
decreasing towards the outer boundary of the heated region.
 Multi-wavelength phase resolved spectroscopy and  more realistic  spectral and light curve models accounting for  non-uniform 
 temperature distributions over the entire surface of the NS and its polar caps are needed to further tune the parameters 
 of the pulsar (see previous paragraph).

Other interesting conclusions follow from the assumption that the second X-ray absorption feature near 0.3 keV is real. 
This and the earlier reported feature at 0.55 keV can represent  consequent  harmonics of cyclotron absorption 
in the spectrum of PSR~B0656+14.
The difference between energies of the lines $\Delta E \approx 0.25$~keV is the cyclotron energy, which allows us to estimate 
the magnetic field in the region where they could be created: $B_{12}= 0.086(1+z)\Delta E (m/Zm_e)$,  where $B_{12}$ is the
magnetic field in $10^{12}$ G units, z is the gravitational redshift 
$1+z = 1 / \sqrt{1-2.953 M_{NS}(M_{\sun})/R_{NS}~(km)}$, 
$Z$ and $m$ are the charge and mass of the plasma particle involved into the absorption, and  $m_e$ is the
electron mass. Assuming electron cyclotron lines 
($m=m_e, Z=1, R_{NS}=13 \mathrm{km}, M_{NS}=1.4 M_{\sun}$), 
$B_{12} \approx 0.25$, and  for the proton lines ($m \approx 1836m_e$),  $B_{12} \approx 45.3$. Both values are about twice
smaller than those  obtained by \citet{2018XMM} assuming that the 0.55 keV feature is the first cyclotron harmonic.  
The canonical dipole magnetic field of \psr\ at the magnetic equator of the NS estimated from the pulsar spin-down, $B_{12}
\approx 4.7$, lies between these estimates. This means that the lines are either created in the pulsar magnetosphere or the
magnetic field at the NS surface is about 5--10 times stronger and differs from the dipole  configuration. 
The latter situation was observed for several pulsars: PSR J0030+0451 \citep{2019ApJ...887L..23B, 2019ApJ...887L..24M,
2019ApJ...887L..21R}; PSR J0108-1431 \citep{2019MNRAS.489.4589A}; PSR J0437-4715 
\citep{2019MNRAS.490.1774L}.

\section{Summary}

Using GTC/OSIRIS and GTC/CIRCE instruments we detected  the pulsar B0656$+$14 in  
the narrow $F657$, $F602$, $F754$,  $F902$ optical bands and in the near-IR $JHK_s$  bands. 
The displacement of the pulsar counterpart in the data obtained in the last twenty years corresponds
to the proper motion that was expected from  the Very Long Baseline Array  radio observations \citep{Brisken:2003aa}.
The  narrow-band radiation fluxes of the pulsar are consistent with those measured earlier in broad 
optical bands. Our  $K_s$ data extends, for the first time, the  optical-near-IR SED of the pulsar 
to the 2.2 $\mu$m range. 
We also present a detailed analysis of the first spectral optical observations of the pulsar obtained by us with the VLT.  
Using the nearest to the pulsar bright field star detected with a high $S/N$ with the GTC and 
\textit{HST} as a secondary photometric standard, we clarified the near-IR fluxes of the pulsar 
obtained with the  \textit{HST} in the   $F110W$, $F160W$ and $F187W$ broad bands. For the consistency 
check, we also carefully reanalysed the \textit{HST} optical photometric data obtained with 
the narrow-band ramp filters $FR459M$, $FR647M$ and $FR914M$. Combining new results   with 
the broad-band optical data obtained earlier with the \textit{BTA}, Subaru and \textit{HST} telescopes, 
we find that the observed optical-near-IR SED of the pulsar provided by the spectral and photometric 
data is well described by a nearly flat  power-law with the spectral index $\alpha \approx 0.6$. 
We do not  see any significant 
narrow spectral features exceeding  $\sim$2$\sigma$
 of standard deviations of the continuum in the pulsar SED. A marginal flux excess over the power-law fit 
 in the $F187W$ and $K_s$ bands is visible  suggesting the pulsar flux increase towards the IR, as was 
 indicated by \citet{2005A&A...440..693S} and \citet{2011ApJ...743...38D}. 
 Higher quality spectral and photometric   data are necessary  to confidently confirm 
 the IR excess.

Using the  nIR-optical SED, the published UV data obtained with the \textit{HST}, and the best archival X-ray 
data 
from \textit{XMM-Newton} and \textit{NuStar}, we performed the self-consistent 
multi-wavelength spectral  analysis  of the rotation phase integrated  spectrum of the pulsar 
from the near-IR through hard X-rays  until 20 keV. We find that the best spectral fit is provided 
by the absorbed combined model consisting of the cool black-body emission from the bulk of 
the surface of the NS, the hot black-body emission from its hot magnetic polar caps, the broken 
power-law emission of the pulsar magnetosphere origin and the absorption line located near 
0.55 keV revealed by  \citet{2018XMM}. 
The temperature of the thermal component  from the
 bulk of the surface of the NS as measured by a distant observer, $kT^{\infty}\approx68$~eV  
 ($T^{\infty}\approx7.9\times10^5$~K), is derived from the multi-wavelength 
 fit  with the accuracy of $\approx$1 eV at the 90\% credibility level. 
 The latter is a factor of four  better 
 than that obtained  using the  X-ray data alone. 
For comparison of this result with NS cooling scenarios see \citet{2020potekhin}.  

 The respective circumferential  
 radius of emission region of $9.3\pm0.4$~km is smaller than a typical radius of an NS,  
 $R_{NS}=13$~km, which suggests nonuniform temperature distribution over the star surface. 
 This is in accord with the results of the rotational phase resolved X-ray spectroscopy 
 by  \citet{2018XMM}. The magnetosphere  nonthermal spectral component of the pulsar  steepens 
 from the optical to X-rays and shows  the spectral break likely located near 0.1 keV.  
 Including the poorly calibrated \textit{XMM-Newton} data below 0.3 keV in our spectral analysis 
 reveals a possible presence of the second absorption 
 line in the spectrum of the pulsar near $\sim$0.3 keV. Assuming that this line and the 0.55 keV line 
 are formed by cyclotron absorption of the thermal emission from the NS surface, this could 
 be used for direct measurements of the pulsar magnetic field. However, the presence of 
 the 0.3 keV line has to be confirmed by better calibration and \textit{eROSITA} observations. 
 
Our multi-wavelength spectral analysis provides much better constraints on the pulsar parameters 
than that based on in a single spectral domain. For instance, it gives much more precise and robust 
location of the pulsar in the $T$--$N_H$ plane shown in  Fig.\ref{FigX-rayFits} as compared to  
previous studies. 
Black-body fits of the thermal spectral components indicate that this NS is probably covered by a bare 
condensed matter.  Application of the bare NS  emission model  assuming a plausible distribution of 
the temperature over the NS surface  and accounting for the pulsar viewing geometry would be 
the next physically reasonable  step in the interpretation of the current and 
future multiwavelength data of \psr. 
This can provide the most robust constraints on the thermal emission parameters 
of the NS  based on the analyses of the phase resolved spectra and light curves   
 in different spectral bands. This will also enable one to measure  $R_{NS}$ and $M_{NS}$ and thus put 
 strong constraints on the EoS of super-high density matter in  interiors of  NSs. The constraints on the pulsar 
 parameters obtained here can be used as initial conditions for such studies.

\label{sec:con}

\section*{Data Availability}

The data underlying this article will be shared on reasonable request to the corresponding author.

\section*{Acknowledgements}
We are grateful to the anonymous referee for the valuable comments which helped to improve this paper.
The manuscript is based on observations made with the Gran Telescopio Canarias (GTC), 
installed in the Spanish Observatorio del Roque de los Muchachos of the Instituto de Astrof\'isica de Canarias, 
in the island of La Palma. 
Development of CIRCE was supported by the University of Florida and the National Science Foundation (grant AST-0352664), 
in collaboration with IUCAA. 
{\sc IRAF} is distributed by the National Optical Astronomy Observatory, which is operated by the Association of 
Universities for Research in 
Astronomy (AURA) under a cooperative agreement with the National Science Foundation. 
This research has made use of the USNOFS Image and Catalogue Archive operated by the United States Naval Observatory, 
Flagstaff Station 
(http://www.nofs.navy.mil/data/fchpix/).
This publication makes use of data products from the Two Micron All Sky Survey, which is a joint project of the 
University of Massachusetts and the Infrared Processing and Analysis Center/California Institute of Technology, 
funded by the National Aeronautics and Space Administration and the National Science Foundation.
Some/all of the data presented in this paper were obtained from the Mikulski Archive for Space Telescopes (MAST). 
STScI is operated by the Association of Universities for Research in Astronomy, Inc., under NASA contract NAS5-26555.
DAZ thanks Pirinem School
of Theoretical Physics for hospitality. R.E.M. gratefully acknowledges support by FONDECYT 1190621, 
and the Chilean Centro de Excelencia en Astrof{\'{i}}sica
y Tecnolog{\'{i}}as Afines (CATA) BASAL grant AFB-170002.
SZ acknowledges PAPIIT grant IN102120.






\bibliographystyle{mnras}
\bibliography{0656}

\bsp	
\label{lastpage}
\end{document}